**Population genomics of sub-Saharan *Drosophila melanogaster*:**
**African diversity and non-African admixture**

John E. Pool[1]*, Russell B. Corbett-Detig[2], Ryuichi P. Sugino[1], Kristian A. Stevens[3], Charis M. Cardeno[3], Marc W. Crepeau[3], Pablo Duchen[4], J. J. Emerson[5], Perot Saelao[3], David J. Begun[3], and Charles H. Langley[3]

[1] Laboratory of Genetics, University of Wisconsin-Madison, Madison, WI, USA

[2] Department of Organismal and Evolutionary Biology, Harvard University, Cambridge, MA, USA

[3] Department of Evolution and Ecology, University of California-Davis, Davis, CA, USA

[4] Section of Evolutionary Biology, University of Munich, Planegg-Martinsried, Germany

[5] Department of Integrative Biology, University of California-Berkeley, Berkeley, CA, USA

* Corresponding author: jpool@wisc.edu

NOTE:

Supplemental figures and tables are not included with this submission, but can be downloaded at: http://www.johnpool.net/Pool2012supplemental.zip

ABSTRACT

*Drosophila melanogaster* has played a pivotal role in the development of modern population genetics. However, many basic questions regarding the demographic and adaptive history of this species remain unresolved. We report the genome sequencing of 139 wild-derived strains of *D. melanogaster*, representing 22 population samples from the sub-Saharan ancestral range of this species, along with one European population. Most genomes were sequenced above 25X depth from haploid embryos.

Results indicated a pervasive influence of non-African admixture in many African populations, motivating the development and application of a novel admixture detection method. Admixture proportions varied among populations, with greater admixture in urban locations. Admixture levels also varied across the genome, with localized peaks and valleys suggestive of a non-neutral introgression process. Genomes from the same location differed starkly in ancestry, suggesting that isolation mechanisms may exist within African populations.

After removing putatively admixed genomic segments, the greatest genetic diversity was observed in southern Africa (*e.g.* Zambia), while diversity in other populations was largely consistent with a geographic expansion from this potentially ancestral region. The European population showed different levels of diversity reduction on each chromosome arm, and some African populations displayed chromosome arm-specific diversity reductions. Inversions in the European sample were associated with strong elevations in diversity across chromosome arms.

Genomic scans were conducted to identify loci that may represent targets of positive selection within an African population, between African populations, and between European and African populations. A disproportionate number of candidate selective sweep regions were located near genes with varied roles in gene regulation. Outliers for Europe-Africa $F_{ST}$ were found to be enriched in genomic regions of locally elevated cosmopolitan admixture, possibly reflecting a role for some of these loci in driving the introgression of non-African alleles into African populations.

AUTHOR SUMMARY

Improvements in DNA sequencing technology have allowed genetic variation to be studied at the level of fully sequenced genomes. We have sequenced more than 100 *Drosophila melanogaster* genomes originating from sub-Saharan Africa, which is thought to contain the ancestral range of this model organism. We found evidence for recent and substantial non-African gene flow into African populations, which may be driven by natural selection. The data also helped to refine our understanding of the species' history, which may have involved a geographic expansion from southern central African (*e.g.* Zambia). Lastly, we identified a large number of genes and functions that may have experienced recent adaptive evolution in one or more populations.

An understanding of genomic variation in ancestral range populations of *D. melanogaster* will improve our ability to make population genetic inferences for worldwide populations. The results presented here should motivate statistical, mathematical, and computational studies to identify evolutionary models that are most compatible with observed data. Finally, the potential signals of natural selection identified here should facilitate detailed follow-up studies on the genetic basis of adaptive evolutionary change.

INTRODUCTION

*Drosophila melanogaster* has a well known history and ongoing role as a model organism in classical and molecular genetics. Its well-annotated genome [1,2] and genetic toolkit have also made it an important model organism in the field of population genetics, in many cases motivating the development of broadly applicable theoretical models and statistical methods. Prior to the advent of DNA sequencing, studies of inversions and allozymes in *D. pseudoobscura* [3,4], and later *D. melanogaster* [5,6], provided some of the field's first glimpses of genetic polymorphisms within and between populations, often providing evidence for geographic clines consistent with local adaptation.

The analysis of DNA sequence data from the *Drosophila Adh* gene motivated the development of methods that compare polymorphism and divergence at different gene regions [7] or functional categories of sites [8], and offered examples of non-neutral evolution. Sequence polymorphism data from additional *D. melanogaster* genes revealed that recombination rate is strongly correlated with nucleotide diversity but not between-species divergence in *D. melanogaster* [9]. This result suggested that genetic hitchhiking [10] could be an important force in molding diversity across the *Drosophila* genome, but it also motivated the suggestion that background selection against linked deleterious variants [11] should likewise reduce diversity in low recombination regions of the genome.

Larger multi-locus data sets initially came from studies of microsatellites and short sequenced loci. Several of these studies compared variation between ancestral range populations from sub-Saharan Africa and more recently founded temperate populations from Europe, finding that non-African variation is far more strongly reduced on the X chromosome than on the autosomes [12,13]. Sequence data also allowed larger-scale comparisons of polymorphism and divergence, leading to suggestions that significant fractions of substitutions at nonsynonymous sites [14] and non-coding sites [15] were driven by positive selection.

Although previous studies have found considerable evidence for a genome-wide influence of natural selection, a thorough and confident identification of recent selective sweeps in the genome requires an appropriate neutral null model that incorporates population history. Both biogeography [16] and genetic variation [17,18] indicate that *D.*

*melanogaster* originated within sub-Saharan Africa. Even within Africa, *D. melanogaster* has only been collected from human-associated habitats, and so its original habitat and ecology, along with the details of its transition to a human commensal species, remain unknown [19]. A few studies have found populations from eastern and southern Africa to be the most genetically diverse [18,20,21], suggesting that the species' ancestral range may lie within these regions. Small but significant levels of genetic structure are present within sub-Saharan Africa [18], which could reflect either long-term restricted migration or short-term effects of bottlenecks associated with geographic expansions within Africa.

On the order of 10,000 years ago [22-24], *D. melanogaster* is thought to have first expanded beyond sub-Saharan Africa, perhaps by traversing formerly wetter parts of the Sahara [16] or the Nile Valley [18]. This expansion involved a significant loss of genetic variation [12,13], brought *D. melanogaster* into the palearctic region (northern Africa, Asia, and Europe), and largely gave rise to the "cosmopolitan" populations that live outside sub-Saharan Africa today. American populations were founded only within the past few hundred years [25], and their complex demography appears to involve admixture between European and African source populations [26].

Recent advances in DNA sequencing technology have allowed genetic variation to be studied on the whole-genome scale. The sequencing of six *D. simulans* genomes [27] provided the first comprehensive look at fluctuations of polymorphism and divergence across the genome and their potential causes, including potential targets of adaptive evolution. More recently, larger samples of *D. melanogaster* genomes have been sequenced, yielding further insight into the potential impact of natural selection on diversity across the *Drosophila* genome [28,29] and connections between genetic and phenotypic variation [29]. However, a large majority of the sequenced genomes are of North American origin, and before we can clearly understand the demographic history of that population, we must investigate genomic variation in its African and European antecedents.

Here, we use whole genome sequencing and population genomic analysis to examine genetic variation in wild-derived population samples of *D. melanogaster*. We use a new method to detect pervasive admixture from cosmopolitan into sub-Saharan populations. We use geographic patterns of genetic diversity and structure to investigate the history of *D. melanogaster* within Africa.  Finally, we identify loci with unusual patterns of allele frequencies within or between populations, which may represent targets of recent directional selection.

RESULTS:

With the ultimate aim of identifying population samples of importance for future population genomic studies, we sequenced genomes from 139 wild-derived *D. melanogaster* fly stocks.  These genomes represented 22 population samples from sub-Saharan Africa and one from Europe (Figure 1; Table S1; Table S2).  Most of these genomes were obtained from haploid embryos [30].  These genomes were found to be essentially homozygous (with the exception of chromosome 2 from GA187 [28]).  A smaller number of genomes were sequenced from homozygous chromosome extraction lines; those included in the published data were found to be homozygous for target chromosome arms (X, 2L, 2R, 3L, 3R).  Three genomes (the ZK sample; Table S1) were sequenced from adult flies from inbred lines; these were found to have extensive residual heterozygosity (results not shown).  The data we analyze below consists entirely of non-heterozygous sequences from haploid embryo genomes.  Apparently heterozygous sites in target chromosome arms were observed at low rates in all genomes, potentially resulting from cryptic copy number variation or recurrent base-calling errors, and were excluded from analysis.  Based on the rarity of such sites (approximately one per 20 kb on average), their exclusion seems unlikely to strongly influence genome-wide summary statistics.

Sequencing was performed using the Illumina Genome Analyzer IIx platform.  Paired-end reads of at least 76 bp were sequenced for each genome (Table S2).  Alignment was performed using BWA [31], with consensus sequences generated via SAMtools [32].  Reads with low mapping scores (<20) were discarded, and positions within 5 bp of a consensus indel were masked (treated as missing data) for the genome in question.  Resequencing of the reference strain $y^1$, $cn^1$, $bw^1$, $sp^1$ and the addition of simulated genetic variation allowed the quality of assemblies to be assessed.  Based on the inferred tradeoff between error rates and genome-wide coverage, a nominal BWA quality score of Q31 (corresponding to an estimated Phred score of Q48) was chosen as a quality threshold for subsequent analyses (Figure S1). Genomic regions with long blocks of identity-by-descent (IBD) consistent with relatedness were masked (Table S3; Table S4). A full description of data generation and initial analysis can be found in the Materials and Methods section. Processed and raw sequence data for all genomes can be found at http://www.dpgp.org/dpgp2/DPGP2.html and http://ncbi.nlm.nih.gov/sra.

*Correlation of sequencing depth with genomic coverage and genetic distance*

The sequenced genomes vary significantly in mean sequencing depth (average number of reads at a given bp) present in the assemblies. Among genomes with relevant data from all five target chromosome arms, mean depth ranges from 18X to 47X (Table S2). Depth was found to have a substantial influence on pairwise genetic distances. Mean depth showed positive, non-linear relationships with distance from the *D. melanogaster* reference genome, and with average distances to other African samples, such as Zambia-Siavonga (Figure 2A). The relationship between depth and genetic distance from Zambia is especially strong (Spearman $\rho$ = 0.63; $P$ < 0.00001), suggesting that population ancestry has little influence on this quantity (a property of the ZI sample further discussed below). This correlation is especially pronounced for genomes with depth below 25X, while only a modest slope is present above this threshold.

Mean depth was also correlated with genomic coverage – the portion of the genome with a called base at the quality threshold (Figure 2B; Spearman $\rho$ = 0.62; $P$ < 0.00001). The lowest depth genomes were found to have ~2% lower coverage than a typical genome with average depth. Some correlation of depth with genetic distance and genomic coverage might be expected if genomes with higher depth were more successful in mapping reads across genomic regions with high levels of substitutional (and perhaps structural) variation. Additionally, a consensus-calling bias in favor of the reference allele, such that higher depth genomes were more likely to have adequate statistical evidence favoring a non-reference allele, might contribute to the reduced genetic distances and genomic coverage exhibited by the lowest depth genomes.

The influence of depth on genetic distance has the potential to bias most population genetic analyses. We found that strict sample coverage thresholds (only analyzing sites covered in most or all assemblies) could ameliorate the depth-distance correlation, but at the cost of excluding most variation and introducing a substantial reference sequence bias (Figure S2). Instead, we addressed the depth-distance issue by focusing most analyses on genomes with >25X depth and made additional corrections when needed, as described below. Assemblies derived from haploid embryos with >25X depth were defined as the "primary core" data set (Table S1). Haploid embryo genomes with <25X depth were denoted as "secondary core". Genomes not derived from haploid embryos were labeled "non-core", and were not analyzed further in this study.

*Identification of cosmopolitan admixture in sub-Saharan genomes*

Previous work has suggested that introgression from cosmopolitan sources (*i.e.* populations outside sub-Saharan Africa) may be an important component of genetic variation for at least some African populations of *D. melanogaster* [18,33,34]. Preliminary examination of this data set revealed a number of sub-Saharan genomes with unusually low genetic distances to cosmopolitan genomes (the latter represented by the European FR sample and the North American reference genome). Undetected admixture could undermine the demographic assumptions of many population genetic methods, altering genetic diversity and population differentiation, and creating long-range linkage disequilibrium. Hence, we attempted to identify specific chromosome intervals that have non-African ancestry, so that they could be filtered from downstream analyses when appropriate.

We developed a Hidden Markov Model (HMM) method to identify chromosome segments from sub-Saharan genomes that have cosmopolitan ancestry, as described under Materials and Methods. The method utilized a "European panel" (the FR sample) and an "African panel" (the RG sample) which may contain some admixture. Because of the diversity-reducing out-of-Africa bottleneck, non-African genomes should be more closely related to each other than they are to African genomes. Therefore, if we examine genomic windows of sufficient length, genetic distances between two FR genomes should be consistently lower than between an RG and an FR genome (Figure S3). To take advantage of this contrast, we constructed chromosome arm-wide emissions distributions by evaluating two locally rescaled quantities in ~50 kb windows. One distribution, representing African ancestry, was formed from genetic distances of each RG genome to the FR panel. The other distribution, representing non-African ancestry, was formed from genetic distances between each FR genome and the remainder of the FR panel. Individual African genomes were then compared to the FR panel to determine the likelihood of African or non-African ancestry in each window (essentially, using the emissions distributions to determine whether we are truly making an Africa-Europe genetic comparison, or if we are actually comparing two non-African alleles in the case of an admixed African genome). The HMM was then applied to convert likelihoods to admixture probabilities for each genome in each window. This approach was validated using simulations (see Materials and Methods; Figure S4). For the empirical data, the above approach was applied iteratively to the RG sample to eliminate non-African intervals from the "African panel" used to create emissions distributions. Emissions distributions generated using the FR and RG samples were also used to calculate admixture probabilities for the other sub-Saharan primary core genomes. Simple correction factors were applied to account for the effects of sequencing depth and other quality factors for each genome (Materials and Methods).

When applied to the RG primary core genomes, the admixture detection method produced generally sharp peaks along chromosome arms, with only 3.3% of window admixture probabilities between 0.05 and 0.95 (Figure S5; Table S5). When primary core genomes from other population samples were analyzed, results still appeared to be of reasonable quality, with 8.3% "intermediate" admixture probabilities as defined above (Figure S5; Table S5). However, inferences for the secondary core and non-core genomes appeared less reliable, with 22.5% intermediate admixture probabilities and more admixture predicted in general (Figure S5; Table S6). Hence, the influence of lower sequencing depth may have added significant "noise" into the admixture analysis. Below, we focus on admixture inferences from the primary core genomes only.

*Inter-population variability in cosmopolitan admixture proportion*

The estimated proportion of cosmopolitan admixture varied dramatically among the twenty sub-Saharan population samples represented in the primary core data set (Figure 3A, Table S7). In general, populations with substantial admixture were observed across sub-Saharan Africa, but admixture proportion varied substantially within geographic regions. At the extremes, one Zambia sample (ZI) had 1.4% inferred admixture among four genomes, while another Zambia sample (ZL) had 84% inferred admixture from the single genome sequenced. A Kruskal-Wallis test for the 14 populations with n $\geq$ 3 primary core genomes supported a significant effect of population on admixture proportion ($P$ < 0.0001).

Testing whether admixture might be related to anthropogenic activity, we found that human population size of the collection locality had a strong positive correlation with admixture proportion (Spearman $\rho$ = 0.60; one-tailed $P$ = 0.003; Figure S6). For the seven collection sites with population sizes below 20,000, all but one population sample had an admixture proportion below 7% (the exception, KR, may reflect a higher regional effect of admixture in Kenya). In contrast, for the eight cities with a population above 39,000, admixture proportion was always above 15%. These results mirror previous findings that urban African flies are genetically intermediate between rural African flies and European flies, when population samples from the Republic of Congo [33,35] and Zimbabwe [34] were examined. Our results suggest that African invasion by cosmopolitan *D. melanogaster* is not limited to the largest African cities, and has occurred in moderately sized towns and cities across sub-Saharan Africa.

In theory, higher admixture levels in urban African locations could result from either neutral or adaptive processes. If larger cities are more connected to international trade, then selectively neutral immigration would affect urban populations first. However, the large size of admixture tracts (*e.g.* a mean admixture tract length of 4.8 centiMorgans or 3.8 Mb for the RG sample; Table S7) suggests an unusually rapid spread of cosmopolitan alleles into Africa, which may not be compatible with plausible levels of passive gene flow. We used the method of Pool and Nielsen [36] to estimate, for the RG sample, the parameters of a two epoch migration rate change model. This method found the highest likelihood for a change in migration 59 generations before present, with near-zero migration before this time (point estimate $1.2 \times 10^{-8}$), and an unscaled migration rate of 0.0010 since the change. It is not clear whether a neutral model invoking thousands of immigrants per generation should be viewed as realistic. Note that the rate of admixture would have to be higher yet if this small Rwandan town was not the point of African introduction for cosmopolitan immigrants.

Alternatively, cosmopolitan admixture into sub-Saharan *D. melanogaster* could be a primarily adaptive phenomenon. Certain cosmopolitan alleles might provide a selective advantage in modern urban environments and may now be favored in modernizing African cities, but may be neutral or deleterious in rural African environments. Or, some cosmopolitan genotypes (such as those conferring insecticide resistance [37]) may now be advantageous in both urban and rural African environments, but have thus far spread primarily into urban areas. In either scenario, there is still a role for demography (*i.e.* migration rates within Africa) in governing the geographic spread of cosmopolitan alleles into African environments in which they are adaptive.

*Intra-population variability in cosmopolitan admixture proportion*

Perhaps more striking than the between-population pattern of admixture are the stark differences in ancestry observed within populations. This individual variability is well-illustrated by the RG sample (Figure 3B), but similar patterns are also observed in other populations (Table S8). Among the 22 RG primary core genomes, nine have no inferred admixture at all, eight others have less than 3% admixture, while the other five genomes contain 20–76% admixture. Based on forward simulations with recombination and migration [36], the observed variance among genomes in cosmopolitan admixture proportion for the RG sample was found to be unlikely under the point estimates of demographic parameters reported above (one-tailed $P$ = 0.02).

The unexpectedly high variance in admixture proportion may require a combination of biological explanations. Inversion frequency differences between African and introgressing chromosomes would reduce the rate of recombination, potentially keeping admixture in longer blocks. However, the genome-wide prevalence of long admixture tracts (including in regions that do not overlap common inversions) makes this explanation incomplete at best.

Alternatively, African populations may be subject to local heterogeneity for any number of environmental factors, and cosmopolitan alleles may confer a greater preference for and/or fitness in specific microhabitats. Such differences might provide a degree of spatial isolation between flies with higher and lower levels of admixture. However, the RG sample was collected from a handful of markets, restaurants, and bars in the center of the relatively small town of Gikongoro, Rwanda (an area less than 200m across), and it's not clear whether any meaningful isolation could exist on this scale.

Finally, sexual selection may play a role in generating this pattern. African strains of *D. melanogaster* are known to display varying degrees of "Z-like" mating behavior, in which females discriminate against males from "M-like" strains, which include cosmopolitan populations [38,39]. Hence, one would expect many African females to avoid mating with males carrying the cosmopolitan alleles responsible for the M phenotype. And indeed, mating choice experiments [33] found that matings between rural Brazzaville females and urban (apparently admixed) Brazzaville males were much less frequent than homogamic pairings. This phenomenon might help to explain the prevalence of admixture in only a subset of genomes in RG and other samples. Further empirical and predictive studies will be needed to assess the ability of these and other hypotheses to explain the inferred patterns of cosmopolitan admixture among sub-Saharan genomes.

*Intra-genomic variability in cosmopolitan admixture proportion*

If cosmopolitan admixture is partly due to adaptive processes, it may be worthwhile to examine variability in admixture proportion across the genome. Figure 4 shows the number of primary core genomes with admixture probability above 50% for each window analyzed by the admixture HMM. By including admixture tracts from 95 sub-Saharan genomes across all populations, we may lose some population-specific signals, but we gain resolution that would not exist within small samples.

Clear differences were observed between chromosomes in admixture levels. Averaging across all windows, arms 3L and 3R had the highest number of admixed genomes (18.1 and 18.0, respectively), while 2L and 2R were somewhat lower (both averaged 14.7).  Both autosomes, however, were considerably more admixed than the X chromosome, which averaged just 9.3 admixed genomes per window.  A qualitatively consistent pattern has been reported [34] in which cosmopolitan admixture was detected on the third chromosome but not the X chromosome in a sample from Harare, Zimbabwe.

      A lesser contribution of the X chromosome to cosmopolitan admixture might be expected if males contributed disproportionately to introgression.  However, the mating preferences described above might be expected to yield the opposite result, suppressing genetic contributions from cosmopolitan males into African populations.  Additionally, the loci responsible for the M/Z behavioral polymorphism are thought to reside primarily on the autosomes [38], which should impede autosomal introgression rather than X-linked introgression.  Another explanation for the deficiency of X-linked admixture is more efficient selection due to the X chromosome's hemizygosity [40].  The X chromosome might have experienced a higher rate of "out-of-Africa" selective sweeps [12], and even though some cosmopolitan adaptations may now be favored in Africa, it is conceivable that the X chromosome contains a greater density of cosmopolitan alleles that are still deleterious in sub-Saharan Africa and limit X-linked introgression.  Even if cosmopolitan alleles that remain deleterious in Africa occur at similar rates on the X chromosome and autosomes, selection might be more effective against introgressing X chromosomes.  Alternatively, the X chromosome's higher recombination rate may lead advantageous X-linked cosmopolitan alleles to introgress within smaller chromosomal blocks.  The recombination rate difference predicted by mapping crosses [28] will be magnified by the lack of recombination in males, and perhaps also by the autosomes' generally higher levels of inversion polymorphism [41], which should decrease autosomal recombination rates in nature and increase X chromosome recombination (due to the interchromosomal effect [42]).

      Considerable variation in the proportion of admixed individuals was also apparent within chromosomes.  For example, the X chromosome's dearth of admixture was most dramatic for the telomere-proximal half of its windows (average 6.9 admixed genomes) and less severe for the centromere-proximal half (average 11.7).  On a finer scale, the proportion of admixed genomes showed relatively narrow genomic peaks and valleys (Figure 4), with the most extreme admixture levels often limited to intervals on the order

of 100 kb.  If the adaptive hypothesis of cosmopolitan admixture is correct, genomic peaks and valleys of admixture could include cosmopolitan loci that are advantageous and deleterious, respectively, in sub-Saharan Africa.  We return to the specific content of these intervals later, in the context of out-of-Africa sweeps.

*Principal Components Analysis*

In order to evaluate the effectiveness of admixture identification and to examine geographic gradients of genetic variation, principal Components Analysis (PCA) [43] was applied to admixture-filtered and unfiltered data.  In both cases, the first principal component clearly reflected cosmopolitan versus African ancestry.  Comparison of these results suggested that our admixture detection method had successfully filtered most, but not all, cosmopolitan admixture from sub-Saharan genomes (Figure 5A).  For example, RG35 was by far the most admixed genome in its Rwanda population sample, with a pre-filtering -PC1 of 0.153.  After filtering, its PC1 dropped to -0.001 – a considerable improvement, although slightly higher than the population average of -0.049.  Hence, a minority of admixture may remain undetected, and for analyses that may be especially sensitive to low levels of admixture, users of the data could opt to exclude genomes with higher levels of detected admixture.

Focusing on PCA from admixture-filtered sub-Saharan data, PC1 separated southern African populations from western African and Ethiopian populations, with eastern African samples having intermediate values (Figure 5B).  PC2 mainly distinguished Ethiopian samples from all others, while subsequent principal components lacked obvious geographic patterns (Table S9).

*Geographic patterns of sub-Saharan genetic diversity*

Nucleotide diversity was evaluated for windows and for full chromosome arms, in terms of both absolute and relative $\pi$ (the latter based on comparison with the RG sample). The use of relative $\pi$ allowed unbiased comparisons of diversity involving populations with incomplete genomic coverage of admixture-filtered data, and it enabled populations lacking two or more genomes with >25X depth to be considered by using RG genomes with similar depth for comparison (see Materials and Methods).

Under simple demographic scenarios of geographic expansion, populations with the highest genetic diversity are the most likely to reflect the geographic origin of all extant populations. Hypotheses for the ancestral range of *D. melanogaster* within sub-Saharan Africa have ranged from western and central Africa (based on biogeography [16]) to eastern and southern Africa (based on a smaller sequence data set [18]). Among 19 African populations, the greatest diversity was found in the ZI sample collected from Siavonga, Zambia (Figure 6; Table 1; Table S10), followed by the geographically proximate ZS and ZO samples. The inferred nucleotide diversity of ZI (0.70%; 0.83% for higher recombination regions) is lower than estimates for geographically similar samples based on multilocus Sanger sequencing [13], and slightly lower than a recent population genomic analysis [28], but higher than an earlier population genomic estimate of $\theta$ [44]. While differences in the genomic coverage of these data sets may help to explain some differences, mapping and consensus-calling biases against non-reference reads may also play a role. Such factors are not expected to have a dramatic impact on comparisons of diversity levels between African populations. Hence, based on the samples represented in our study, southern-central Africa appears to contain the center of genetic diversity for *D. melanogaster*. Although this hypothesis requires further confirmation, these results are consistent with a southern African origin for *D. melanogaster*.

Much of Zambia and Zimbabwe is characterized by a subtropical climate and seasonally dry Miombo and Mopane woodland. Whether this landscape might reflect the original environment of *D. melanogaster* is unclear, because the species has never been collected from a completely wild environment [16] and the details of its transition to an obligately human-commensal species are unknown [19]. Compared with related species, African strains of *D. melanogaster* have superior resistance to desiccation [45] and temperature extremes [46]. These characteristics would be predicted by an evolutionary origin in subtropical southern Africa, as opposed to humid equatorial forests.

Most populations from eastern Africa (including Kenya, Rwanda, and Uganda) had modestly lower diversity compared to Zambia and Zimbabwe, while western populations (including Cameroon, Guinea, and Nigeria) showed an additional slight reduction.  The two Ethiopian samples showed the lowest variation among African populations, with roughly three quarters the diversity of ZI, potentially indicating a bottleneck during or since the species' occupation of Ethiopia.  The distinctness of Ethiopian samples was also indicated by an analysis of mitochondrial and Wolbachia genomes from these same genomes [47].  Otherwise, only two population samples had reduced diversity relative to the overall geographic pattern described above, and one of these (CK) had limited pairwise comparisons in the admixture-filtered data.

The other sample with locally reduced variation, TZ, displays an unusual pattern of diversity loss on chromosome arms X and 2L specifically (Table 1), associated with all three sampled genomes carrying inversions *In(1)A* and *In(2L)t* [48].  Similarly, two of the three Kenya samples (KN and KR) show reduced diversity on arm 2L only, also apparently in association with *In(2L)t* [48].  These results suggest the possibility that selection on polymorphic inversions, which are common in sub-Saharan populations [41], can be an important determinant of genome-scale diversity levels.  Although this hypothesis is contrary to some theoretical predictions [49] and empirical findings [50] that would lead one to expect the effects of inversions to be mainly restricted to breakpoint regions, it is supported by an analysis of inversion polymorphism and linked variation in the genomes studied here [48].

*Chromosomal diversity in a non-African population*

Consistent with previous work [12,13,51], variation for the cosmopolitan sample (FR) is much more strongly reduced on the X chromosome relative to the autosomes (Table 1).  However, further genomic patterns in the ratio of $\pi_{FR}$ to $\pi_{RG}$ can be observed (Figure 7).  This diversity ratio ranges from below 0.2 (at the X telomere) to above 1 (for a window on arm 3R), with similar patterns observed if $\pi_{FR}$ is instead compared against $\pi_{ZI}$ (Figure S7).  Diversity ratios on autosomal arms showed distinct differences:  FR retains 59% of the RG diversity level on arm 2R but 84% on arm 3R (Table 1).

Based on the inversions identified for each genome [48], we examined the influence of inversions on chromosome arm-wide diversity by recalculating $\pi_{FR}$ and $\pi_{RG}$ using standard chromosome arms only. For the RG sample, the exclusion of inversion-carrying arms had negligible influence on diversity, except that the inclusion of *In(3R)P* (present in four of 22 genomes) increased $\pi_{RG}$ on arm 3R by 4% (Table S10). More dramatic contrasts were observed for the FR sample, in which inversions were found to result in arm-wide diversity increases of 10% on arm 2L (due to one of eight FR genomes carrying *In(2L)t*) and 18% on arm 3L (due to a pair of *In(3L)P* chromosomes). As further detailed in a separate analysis [50], arm 3R was even more strongly affected, with a 29% diversity increase due to the presence of *In(3R)P* (in three of eight genomes), *In(3R)K* and *In(3R)Mo* (one genome each). Although the French sample only contains inversions on these three arms, they contribute to a 12% genome-wide increase in nucleotide diversity.

In light of the above observations, it is possible that inversions have had important effects both in reducing chromosome arm-wide diversity (for the Tanzania and Kenya populations) and also in elevating it (for non-African autosomes). As further suggested in a separate analysis [48], the spatial scale of increased diversity associated with inversions in the France sample (Figure 7) may indicate a recent arrival of inverted chromosomes from one or more genetically differentiated populations. Given that similar levels of gene flow are not indicated by polymorphism on chromosome arms lacking inversions, the spread of genetically divergent inverted chromosomes into France may have been primarily driven by natural selection. In light of their powerful elevation of $\pi_{FR}$, inverted chromosomes in this sample may have originated from a more genetically diverse African or African-admixed population. Similarly, the more modest elevation of $\pi_{RG}$ associated with *In(3R)P* might indicate the recent introgression of these inverted chromosomes from a genetically differentiated population. However, the nature of selective pressures acting on inversions in natural populations of *D. melanogaster* remains largely unknown.

Without inversions, relative $\pi_{FR}$ for autosomal arms ranged from 0.58 to 0.63, with chromosome 3 showing higher values than chromosome 2. In light of the above hypothesis to account for the presence of divergent inverted chromosomes in the France sample, some of the remaining differences in relative $\pi_{FR}$ among inversion-free chromosomes might stem from recombination between standard chromosomes and earlier waves of introgressing inverted chromosomes. Alternatively, given that *D. melanogaster* autosomes frequently carry recessive deleterious mutations [52], associative overdominance during the out-of-Africa bottleneck might have favored intermediate inversion frequencies [53-55]. This hypothesis is mainly plausible in small populations [56], which may have existed due to strong founder events during the out-of-Africa expansion [57]. Given the opportunity for recombination between standard and inverted chromosomes since that time, past associative overdominance related to inversions (or centromeric regions) might contribute to the modest difference in relative $\pi_{FR}$ between inversion-free second and third chromosomes, as well as the larger gap between both autosomes and the X chromosome.

The ratio of relative $\pi_{FR}$ for the X chromosome versus the inversion-free autosomes appears consistent with some previously explored founder event models [57] if chromosomes X and 2 are compared (ratio = 0.692, compared to a minimum of 0.669 in the cited study), but not if chromosome 3 is examined instead (ratio = 0.646). Some studies have concluded that the difference between X-linked and autosomal diversity reductions in cosmopolitan *D. melanogaster* exceeds the predictions of demographic models involving population bottlenecks and/or a shift in sex-specific variance in reproductive success [12,13]. Instead, the X chromosome's disproportionate diversity reduction might result from more efficient positive selection on this chromosome (due to male hemizygosity [40]) during the adaptation of cosmopolitan populations to temperate environments. However, it appears relevant that the above studies examined autosomal loci on chromosome 3, but not chromosome 2. Further theoretical, simulation, and inferential studies to elucidate the relative influence of selection, demography, and inversions on the X chromosome and autosomes is needed before their relative contribution to diversity in cosmopolitan *D. melanogaster* can be clearly understood.

*Genetic structure and expansion history*

Levels of genetic differentiation between populations were evaluated in terms of $D_{xy}$ and $F_{ST}$ [58] for each chromosome arm. In order to minimize the effects of any residual admixture in the filtered data, only genomes with admixture proportion below 15% were included. Populations with sufficient data for this analysis included CO, ED, FR, GA, GU, KR, NG, RG, TZ, UG, ZI, and ZS. Within Africa, $F_{ST}$ values on the order of 0.05 were typical (Table 2). Geographically proximate population pairs often had lower $F_{ST}$ (at minimum, a value of 0.009 between ZI and ZS). Comparisons involving the ED sample gave uniformly higher $F_{ST}$ than other African comparisons (median 0.147), consistent with the loss of diversity observed for Ethiopian samples. As expected, comparisons of African samples with the European FR sample yielded the highest $F_{ST}$ values (median 0.208). Genetic differentiation at putatively unconstrained short intron sites [59,60] showed similar patterns (Table S11), but as expected, magnitudes of $D_{xy}$ and $\pi$ were more than twice as high as for all non-centromeric, non-telomeric sites (for ZI, short intron $\pi$ = 0.0194).

In order to assess the compatibility our data with a model of geographic expansion from southern Africa, we examined the ratio of each population's $D_{ZI}$ (average pairwise genetic distance, or $D_{xy}$, between this population and the ZI sample) and $\pi_{ZI}$. This ratio will be near 1 if a population's genomes are no more divergent from ZI genomes than ZI genomes are from each other, consistent with the recent sampling of this population's diversity from a ZI-like ancestral population. In contrast, ratios exceeding 1 indicate that a population contains unique genetic diversity not present in ZI. Populations from eastern Africa (KR, RG, TZ, UG) and Europe (FR) had ratios compatible with a recent ZI-like origin (Table 2). However, populations from western Africa (CO, GA, GU, NG) and Ethiopia (ED) showed modest levels of unique variation. The highest ratio, for Guinea (GU), indicated a 2.9% excess of $D_{ZI}$ over $\pi_{ZI}$. Elevated ratios could indicate a relatively ancient occupation of at least some of the above regions (perhaps on the order of tens of thousands of years). Alternatively, under the hypothesis of an expansion from southern Africa, these regions may have received a genetic contribution from a different part of a structured southern African ancestral range (*e.g.* migration into Gabon and western Africa from Angola, which also contains Miombo woodlands but has not been sampled).

Examination of genomewide genetic differentiation may also shed light on the sub-Saharan origins of cosmopolitan *D. melanogaster*. Geographic hypotheses for expansion of *D. melanogaster* from sub-Saharan Africa have ranged from a Nile route starting from the equatorial rift zone [18] to a more western crossing of the Sahara via formerly wetter areas of "Paleochad" [16]. A simple prediction is that the sub-Saharan samples most closely related to the cosmopolitan source population should show the lowest values of $D_{xy}$ and $F_{ST}$ relative to the cosmopolitan FR sample. However, even low levels of undetected cosmopolitan admixture in sub-Saharan genomes could obscure this signal, and so only genomes with <15% detected admixture were considered below. Among the eleven African populations analyzed (see above), the Kenyan KR sample showed the lowest genome-wide $D_{FR}$ (Table 2) and would have had the lowest FR $F_{ST}$ if not for its anomalous pattern of variation for arm 2L (Table S11). However, KR is the sample with the highest proportion of detected cosmopolitan admixture, which clouds the interpretation of these results. After KR, the lowest $D_{FR}$ values come from the western group of samples (NG, CO, GA, and GU), of which two (CO and GU) had relatively low levels of detected admixture. Despite its northeast sub-Saharan location, the Ethiopian ED sample does not appear to represent a genetic intermediate between cosmopolitan and other sub-Saharan populations, and may instead represent a separate branch of this species' geographic expansion. Further sampling and analysis may be needed to obtain compelling evidence regarding the geographic origin of cosmopolitan *D. melanogaster*.

One scenario for the sub-Saharan expansion of *D. melanogaster* is illustrated by the geographic fit of a simple neighbor-joining population tree based on $D_{xy}$ values (Figure 8; Figure S8). This tree is consistent with the hypothesis of a southern Africa origin for *D. melanogaster*, with an initial expansion into eastern Africa, followed by offshoots reaching Ethiopia, the palearctic (northern Africa and beyond), and western Africa. Of course, even after the filtering of cosmopolitan admixture, a tree-like topology is not likely to fully describe the history of sub-Saharan *D. melanogaster* populations. However, the history described above seems consistent with levels and patterns of population diversity (Figure 6; Table 2), and may capture some important general features of the species' history.

Even if the general expansion history described above ultimately proves to be accurate, many historical details await clarification. Diversity differences among African populations could indicate population bottlenecks during a sub-Saharan range expansion, and population growth during such an expansion is also possible. Further analysis of population genomic data is also needed to establish whether ancestral range populations have also been affected by population growth [23] or a bottleneck [21]. Lastly, although migration within Africa has not erased the observed diversity differences and genetic structure, the historical and present magnitudes of such gene flow are not clear. The quantitative estimation of historical parameters may be addressed by detailed follow-up studies. However, for a species like *D. melanogaster*, in which very large population sizes may allow relatively high rates of advantageous mutation and efficient positive selection, one concern is that the effects of recurrent hitchhiking may be important on a genomewide scale [27,28,61,62]. Hence, the application of standard demographic inference methods to random portions of the *D. melanogaster* genome (or even putatively unconstrained sites) may yield estimates that are biased by violations of the assumption of selective neutrality. Under the assumption of demographic equilibrium, Jensen *et al.* [62] estimated a ~50% reduction in diversity due to positive selection for Zimbabwe *D. melanogaster*, and selective sweeps may have similarly important influences on the means and variances of other population genetic statistics as well. Hence, further methodological development may be needed before accurate demographic estimates can be obtained for species in which large population sizes facilitate efficient natural selection.

*Influence of recombination and selection on genetic variation*

Focusing on our largest population sample (22 primary core RG genomes), we investigated relationships between genetic diversity and mapping-based recombination rate estimates [28]. To minimize the effects of direct selective constraint on the sites examined, we focused on the middles of short introns (bp 8 to 30 of introns ≤65 bp in length), which are among the most polymorphic and divergent sites observed in the *Drosophila* genome [59,60]. Since each 23 bp intronic locus is too small to be considered individually, we show broad-scale patterns of diversity from all relevant sites within a given cytological band. Consistent with previous findings [9], strong relationships between recombination and variation were observed for all chromosome arms (Figure 9), with Pearson's *r* ranging from 0.68 (for 3L) to 0.95 (for 2R), with *P* = 0.0005 or lower for all arms (Table S12). Curiously, bp position along the chromosome arm was a stronger predictor of diversity than estimated recombination rate for arms 3L and 3R (Table S12), which could reflect imprecision in recombination rate estimates for chromosome 3, or the influence of polymorphic inversions on recombination in nature. Across all autosomal arms, the strongest correlation between recombination and diversity was for low rates of crossing-over (adjusted rate below 1 cM/Mb, equivalent to an unadjusted 2 cM/Mb rate,

Pearson $r$ = 0.56 and $P$ = 0.0002). However, a strong correlation persisted above this threshold as well (Pearson $r$ = 0.44, $P$ = 0.002). Correlations within these categories were not significant for the X chromosome, potentially due to smaller numbers of chromosome bands, especially for the low recombination category ($n$=4). Overall, the above results are consistent with the well-supported role for natural selection in reducing variation in regions of low recombination. However, the relative contributions of specific selection models such as hitchhiking [10] and background selection [11] to this pattern have not been quantitatively estimated.

    Examining the RG sample's allele frequencies at short intron sites, we observed an excess of singleton polymorphisms (sites with a minor allele count of 1) for all chromosome arms relative to the predictions of selective neutrality and demographic equilibrium (Figure 10A). The degree of this excess varied among chromosome arms: compared to a null expectation of 31% singleton variants, the autosomal arms ranged from 33% to 37%, while the X chromosome had 44% singletons. The general excess of rare alleles could reflect population growth, as suggested for a Zimbabwe population sample [23], and growth has some potential to influence X-linked and autosomal variation differently [63]. Recurrent hitchhiking may contribute to the genomewide excess of rare alleles [64]. Under this hypothesis, the difference in singleton excess between the X chromosome could reflect more efficient X-linked selection due to hemizygosity [40]. Without a difference in the rate of X-linked and autosomal adaptation, this contrast could instead result from a greater fraction of X-linked selective sweeps acting on new beneficial mutations, with relatively more autosomal sweeps via selection on standing variation. The autosomes may have more potential to harbor recessive and previously deleterious functional variants, and sweeps from standing variation do not strongly influence the allele frequency spectrum [65].

    We also used short intron allele frequencies to conduct a preliminary analysis of the relationship between recombination and rare alleles. Specifically, we tested whether the proportion of singletons among variable sites differed between low recombination regions (defined here as < 1 cM/Mb) and moderate to high recombination regions (> 1 cM/Mb). No clear relationship between recombination and allele frequency was observed above this cutoff (results not shown). For the 1 cM/Mb threshold, the X chromosome showed an

elevated proportion of singletons in the low recombination category (53% *vs.* 43%; Pearson $\chi^2$ *P* = 0.032; Figure 10B). Data from the autosomes are inconclusive: while three arms show non-significant trends toward more rare alleles in low recombination regions (Figure 10B), arm 2L showed a significant pattern in the opposite direction (30% *vs.* 36%; *P* = 0.025), possibly reflecting specific evolutionary dynamics of the 2L centromere-proximal region. The X chromosome result is qualitatively consistent with the predictions of the recurrent hitchhiking model [64] and some (but not all) previous findings from *D. melanogaster* [51, 66, 67]. Under this hypothesis, the lack of a comparable autosomal pattern might indicate a lesser influence of classic selective sweeps on the autosomes relative to the X chromosome, or a greater effect of inversion-related selection on the autosomes obscuring predictions of the recurrent hitchhiking model. Background selection may also increase the proportion of singletons [68,69], although a greater X-linked effect of background selection has not been suggested. Further study is needed to quantify the influence of positive and negative selection at linked sites on nucleotide diversity and allele frequencies in the *D. melanogaster* genome.

*Linkage disequilibrium and its direction*

Linkage disequilibrium (LD) was examined using a standard correlation coefficient ($r^2$) between single nucleotide polymorphism (SNP) pairs, and also via the directional LD metric $r_\omega$ [70, 71]. The $r_\omega$ statistic is positive when minor frequency alleles at two sites tend to occur on the same haplotype, negative if they tend to be on different haplotypes. Although we lack a comprehensive understanding of the evolutionary forces capable of influencing $r_\omega$, it is known that hitchhiking strengthens positive $r_\omega$ (since recombination near a selective sweep leaves groups of positively linked SNPs [72]), while negative $r_\omega$ may result from epistatic interactions among beneficial or deleterious alleles [70]. Empirical data from the RG sample was compared against neutral simulations with equilibrium demographic history. Importantly, equilibrium may not accurately reflect the history of the RG sample: recent population growth may have occurred, and the RG sample's modest diversity reduction compared to the ZI sample may imply a mild population bottleneck. Although the full effects of demography can not be eliminated by any simple procedure, we can reduce the influence of growth or other forces responsible for this population's excess of singleton polymorphisms by excluding singletons from the empirical and simulated data.

In general, an excess of LD was observed over neutral, equilibrium predictions for all chromosome arms (Figure 11A). The X chromosome's lower LD is consistent with its higher average recombination rate (54% higher for the regions examined [28]). The RG pattern contrasts with data from a North American population, which showed elevated X-linked LD [28, 29] that likely reflects a stronger influence of demography and possibly selection on the X chromosome during the species' out-of-Africa expansion. For the RG sample, the X chromosome's LD excess was largely confined to the 10-100 bp scale. In contrast, autosomal arms showed an excess of LD at all scales 10 bp and above (Figure 11A). Since the simulations account for differences in average (inversion-free) recombination rate between the X and autosomes, the autosomes' more pronounced LD excess could result from a stronger influence of inversions on these arms. As noted above, the autosomes' higher inversion polymorphism should reduce autosomal recombination rates in nature and increase X chromosome recombination rates. Arm 3R contains the largest number of common inversions in Africa [41], and LD for this arm is by far the highest. Arm 3R's somewhat lower average recombination rate (7-27% lower than other autosomal arms for the analyzed regions) may contribute to this pattern as well. The above observations regarding LD are concordant with estimates of the population recombination rate for the RG sample, which are elevated for the X chromosome (in spite of its potentially lower population size) and reduced for 3R [73].

Notably, the observed LD excess is driven entirely by SNP pairs with positive $r_\omega$ (Figure 11B), while negative SNP pairs show no departure from equilibrium expectations (Figure 11C). Although cosmopolitan admixture has been largely removed from the analyzed data set, it remains possible that demographic events of this nature might inflate positive LD specifically. Inversions may well play a role in boosting positive LD, since inversion-associated polymorphisms may often be present at similar frequencies on the same haplotypes. However, given the excess of LD on all chromosome arms and on relatively short spatial scales, it is not yet clear whether inversions are a sufficient explanation. Recurrent hitchhiking may also contribute to the genome-wide excess of positive LD [72]. Further studies will be needed to evaluate the compatibility of specific hypotheses with genome-wide LD patterns.

*Potential targets of selective sweeps in a Rwanda sample*

Identifying the genes and mutations underlying Darwinian selection is an important aspect of evolutionary biology, and of population genomics in particular. The lack of a precise demographic model limits our ability to formally reject the null hypothesis of neutral evolution for specific loci, since certain demographic models can mimic the effects of selective sweeps [74]. However, we have still sought to learn about general patterns of directional selection in the genome by conducting a series of local outlier analyses to detect unusual patterns of allele frequencies within and between populations that are consistent with recent adaptive evolution. These outlier analyses necessarily involve a strong assumption about the proportion of the genome affected by selection. However, the enrichment analyses we perform on these outliers should be robust to some level of random false positives within the outliers, and should still be informative if not all adaptive loci are detected.

We searched for putative signals of selective sweeps in the RG sample using a modified version of the SweepFinder program [75,76] that looks for both allele frequency spectra and diversity reductions consistent with recent selective sweeps. As further described in the Materials and Methods section, we analyzed the RG data in windows and used the $\Lambda_{max}$ statistic in an outlier framework, rather than making an explicit assumption regarding the appropriate demographic null model – as would have been required for typical simulations defining statistical significance. Here, we focus on the most extreme 5% of windows from each chromosome arm. After merging neighboring outlier windows, a total of 343 outlier regions were obtained (Table S13). For each outlier region, the gene with the closest exon to the $\Lambda_{max}$ peak was recorded. Genes within extreme outlier regions included *Ankyrin 2* (cytoskeleton, axon extension), *Girdin* (actin filament organization, regulation of cell size), *Laminin A* (behavior, development, meiosis), *narrow abdomen* (ion channel, circadian rhythm), *Odorant receptor 22a* [77], and ribosomal proteins *S2* and *S14b* (separate regions). Several strong outliers corresponded to genes also implicated in a recent genome scan based on outliers for low polymorphism relative to divergence [28], including *bendless* (axonogenesis, flight behavior) *CENP-meta* (mitotic spindle organization, neurogenesis), *female sterile (1) homeotic* (regulation of transcription), *Heterogeneous nuclear ribonucleoprotein at 27C* (regulation of splicing), *loquacious* (RNA interference, nervous system development, germ-line stem cell division), and *no distributive junction* (meiotic chromosome segregation).

Despite a similar number of outlier regions as the $F_{ST}$ analyses described below, the $\Lambda_{max}$ scan yielded a much larger number of significantly enriched gene ontology categories: 115 categories had $P < 0.05$ based on random permutation of target windows within chromosomal arms (Table 3; Table S14). Consistent with previous results from a population genomic outlier analysis of diversity and divergence [28], numerous biological processes related to gene regulation were observed, including positive and negative regulation of transcription, positive regulation of translation, regulation of alternative splicing, mRNA cleavage, chromatin organization, regulation of chromatin silencing, and gene silencing. Many enriched cellular components (*e.g.* nucleus, precatalytic spliceosome, mRNA cleavage and polyadenylation complex, ribonucleoprotein complex, heterochromatin, and euchromatin) and molecular activities (*e.g.* DNA binding, mRNA binding and especially mRNA 3'-UTR binding) were also consistent with a broad importance for regulators of gene expression in recent adaptive evolution. A number of the GO terms listed in Table 3 were also reported from the above-mentioned genome scan [28], including negative regulation of transcription, positive regulation of translation, ribonucleoprotein complex, precatalytic spliceosome, protein ubiquitination, nuclear pore, lipid particle, and spermatogenesis. Other enriched biological processes included oogenesis, neurogenesis, male meiosis and female meiosis chromosome segregation, regulation of mitosis and apoptosis, and phagocytosis. Additional cellular components included microtubule-associated complex, kinetochore, and fusome while enriched molecular activities also included ATP binding and voltage-gated calcium channel activity.

*Locally elevated genetic differentiation between African populations*

Nine African population samples with larger sample sizes after admixture filtering were included in an analysis of local genetic differentiation. $F_{ST}$ was evaluated for each pair of populations, and the mean $F_{ST}$ for each window was noted. Examination of the 2.5% highest mean $F_{ST}$ values for each chromosome arm and the merging of neighboring outlier windows resulted in 294 outlier regions (Table S15). For each outlier region, the gene with the closest exon to the center of the most extreme window was noted. Genes associated with unusually strong $F_{ST}$ outlier regions included *Odorant receptor 22b* (tandem paralog of the above-mentioned *Or22a*), *Cuticular protein 65Au*, *Dystrophin, P-element somatic inhibitor*, and *CG15696* (predicted homeobox transcription factor). Of course, many of the strongest putative signals of adaptive differentiation are wide, and further investigation will be needed to confirm specific targets of selection. Permutation of putative target windows indicated that genes from 34 GO categories were significantly over-represented among our outliers at the $P$ = 0.05 level (Table 4; Table S16). These GO categories included biological processes (*e.g.* oocyte cytoskeleton organization, regulation of alternative splicing, regulation of adult cuticle pigmentation), cellular components (*e.g.* mitochondrial matrix, dendrite), and molecular functions (*e.g.* olfactory receptor activity, mRNA binding).

*Locally elevated genetic differentiation between Africa and Europe*

A windowed $F_{ST}$ outlier approach was also applied to detect loci that may contain adaptive differences between sub-Saharan (RG) and European (FR) populations. Some of these loci might have had adaptive importance during the expansion of *D. melanogaster* into temperate environments, but others could reflect recent selection within Africa. A total of 346 outlier regions resulted from analyzing the upper 2.5% tail of Rwanda-France $F_{ST}$ (Table S17). Genes associated with strong $F_{ST}$ outliers included *Or22a* (which may be under selection in Africa, see above), *CHKov1* (insecticide and viral resistance [78, 79]), *ACXC* (spermatogenesis), and *Jonah 98Ciii* (digestion), plus a number of genes involved in morphological and/or nervous system development (*e.g. Bar-H1*, *Death-associated protein kinase related*, *Enhancer of split*, *hemipterous*, *highwire*, *mastermind*, *rictor*, *sevenless*, *Serendipity δ*, and *wing blister*). Other genes at the center of strong outlier regions were also detected by a genome-wide analysis of diversity ratio between U.S. and Malawi populations [28], including *dpr13* (predicted chemosensory function), *Neuropeptide Y receptor-like*, *rugose* (eye development), and *Sno oncogene* (growth factor signaling, neuron development).

The genes identified in this analysis still yielded 31 significantly enriched GO categories (Table 5; Table S18). Biological processes among these GO categories included chromosome segregation, locomotion, female germ-line cyst formation, histone phosphorylation, and alcohol metabolism. Cellular components included basal lamina and polytene chromosome interband, while molecular activities included transcription coactivators and neuropeptide receptors. The detected GO categories were essentially distinct from those obtained from the diversity ratio analysis of Langley *et al.* (28). The lack of overlap may stem at least partially from differences in the statistics and populations used in each analysis. The well-known challenges of identifying positive selection in the presence of bottlenecks [74], along with uncertainty regarding the portion of the genome affected by adaptive population differences, may also contribute to these findings. Both analyses, however, should motivate new adaptive hypotheses to be tested via detailed population genetic analyses and experimental approaches.

If the rapid introgression of non-African genotypes into African populations documented above is driven by natural selection, then sharp peaks and valleys of admixture along the genome (Figure 4) should contain functional differences between sub-Saharan and cosmopolitan populations. Such differences may have been driven by natural selection after these populations diverged, and hence may be detectable by the Africa-Europe $F_{ST}$ outlier scan presented above. Given that the scale of these $F_{ST}$ outliers (on the order of 10 kb) is narrower than our admixture peaks and valleys (on the order of 100 kb), population genetic signals of elevated differentiation may be helpful in localizing genes responsible for driving or opposing non-African gene flow into African populations.

We selected eight clear genomic peaks of admixture within the higher recombination regions analyzed for $F_{ST}$. These peaks were delimited by windows containing the local maximum number of admixed genomes, and identified $F_{ST}$ outlier regions that either overlapped them or were within 100 kb. Valleys of admixture were more difficult to clearly distinguish from gaps between peaks and minor fluctuations (Figure 4) – three were identified, one of which overlapped several $F_{ST}$ outlier regions (Table S19). For peaks of admixture, seven of these eight regions were associated with $F_{ST}$ outlier regions (Table S19), exceeding random expectations (permutation $P$ = 0.017). Stronger outlier regions associated with admixture peaks included the genes *Bar-H1*, *Enhancer of split*, *Neuropeptide Y receptor-like*, and *sevenless*. Further studies will be needed to evaluate the possibility that cosmopolitan alleles at one or more of these loci may now confer a fitness advantage in urban African environments.

CONCLUSIONS AND PROSPECTS:

    Here, we have described variation across more than one hundred *D. melanogaster* genomes, focusing on the species' sub-Saharan ancestral range. We observed clear evidence of cosmopolitan admixture at varying levels in all sub-Saharan populations. While admixture initially appeared to be merely a barrier to studying African variation, inferred patterns of admixture suggested that this process is associated with intriguing biological dynamics. Based on the apparent speed of introgression, the association of admixture with urban environments, and dramatically differing admixture levels across the genome (with peak admixture levels correlated with outliers for Africa-Europe $F_{ST}$), it appears that admixture may be a primarily non-neutral process. Unexpected variance in admixture proportion within populations provides another departure from simple models, and could indicate isolation mechanisms within African populations.

    We observed the greatest genetic diversity in a Zambian sample and nearby populations, suggesting a possible geographic origin for the species. Even at a broad genomic scale, however, it appears that genetic diversity does not always reflect demographic expectations. We observed chromosome arm-specific deviations in population diversity ratios, most notably for comparisons involving the European population: genetically differentiated inverted chromosomes strongly influence autosomal diversity in our France sample, potentially due to recent natural selection elevating the frequency of introgressing inversions with African origin. Considering this hypothesis alongside our admixture inferences, it is conceivable that selection has driven gene flow in both directions across the sub-Saharan / cosmopolitan genetic divide, with consequences

for genome-wide levels and patterns of diversity.  Additional studies are needed to evaluate models of population history, natural selection, and inversion polymorphism that may account for the above patterns.

We have identified numerous genes and processes that may represent targets of positive selection within and between populations.  However, further investigations will be needed to confirm targets of selection and their functional significance.  Such studies may help reveal the biological basis of this species' adaptation to temperate environments, as well as contrasting environments within Africa, while potentially also providing more general insights into the genetic basis of adaptive evolution.

Although the aims of this publication are primarily descriptive, data such as that presented here may play an important role in resolving some long-standing controversies in population genetics.  It's clear that natural selection plays an important role in shaping sequence divergence between *Drosophila* species and in reducing polymorphism in genomic regions of low recombination.  However, the relative importance of natural selection and neutral forces in governing levels and patterns of variation in regions of higher recombination is unresolved.  We still do not know if, for example, linked hitchhiking events have an important influence on diversity at most sites in the genome.  The relative impact of population history and natural selection on genetic diversity during the out-of-Africa expansion of *D. melanogaster* is also uncertain.  And in regions of low recombination, the relative contributions of hitchhiking and background selection in reducing genetic variation have not been quantified.  It is our hope that population genomic data sets like this one will motivate theoretical and simulation studies that advance our fundamental understanding of how evolutionary forces shape genetic variation.

MATERIALS AND METHODS:

*Drosophila stocks and DNA preparation*

       Genomes reported here are derived from the population samples listed in Table S1 and depicted in Figure 1. The collection methods for samples collected in 2004 or later correspond to a published protocol [80]. Information about individual fly stocks is presented in Table S2. Most of the relevant stocks are isofemale lines, each founded from a single wild-caught female. In some cases, intentional inbreeding was conducted by sib-mating for five generations; such lines have an 'N' appended to the isofemale line label. Although not a focus of our analysis, we have also released genomic data from a small number of chromosome extraction lines, created using balancer stocks.

       Except for the three ZK genomes, DNA for all inbred and isofemale lines was obtained from haploid embryos [30]. Briefly, a female fly from the stock of interest was mated to a male homozygous for the *ms(3)K81$^1$* allele [81]. This mating produces some eggs which are fertilized but fail to develop because the clastogenic paternal genome. Rarely, such eggs bypass apparent checkpoints and develop as haploid embryos. Eggs with partially developed first instars were visually identified under a microscope. DNA was isolated from haploid embryos and genome-amplified as previously described [30]. For the ZK genomes and chromosome extraction lines, DNA was isolated from 30 adult flies (generally females; mixed sexes in the case of autosomal extraction lines). For all samples, library preparation for sequencing (ligation of paired end adapters, selection of ~400bp fragments, and PCR enrichment) was conducted as previously described [30]. In some cases, bar code tags (6 bp) were added to allow multiplexing of two or more genomes in one flow cell lane.

*Sequencing, assembly, and data filtering*

       Sequencing was performed using standard protocols for the Illumina Genome Analyzer IIx. Initial data processing and quality analysis was performed using the standard Illumina pipeline. Sequence reads were deposited in the NIH Short Read Archive as project SRP005599. Alignments to the *D. melanogaster* reference genome (BDGP release 5) using BWA version 0.59 [31] with default settings and the "-I" flag. Program defaults included a

32 bp seed length; reads could therefore map to the reference only if two or fewer reference differences were present within a seed. Although read lengths varied from 76 bp to 146 bp within this data set, only the first 76 bp of longer reads was used for the assemblies reported here. In order to exclude ambiguously mapping reads, those with a BWA mapping quality score less than 20 were eliminated from the assemblies.

Consensus sequences for each assembly were obtained using the SAMtools (version 0.1.16) pileup module [32]. These diploid consensus sequences generally included a few thousand heterozygous calls, scattered across the genome. Such sites are not expected to represent genuine heterozyosity in these haploid/homozygous samples (with the exception of ZK, in which large-scale heterozygosity was observed, presumbaly due to incomplete inbreeding). All putatively heterozygous sites were masked to 'N'. Sites within 5 bp of a consensus indel were also masked to 'N' – this criterion was found to reduce errors associated with indel alignment; no appreciable benefit was observed if 10 bp was masked instead (data not shown).

Data were only considered for "target" chromosome arms, as defined in Table S1. These are chromosome arms expected to derive from the population sample of interest (as opposed to originating from laboratory balancer stocks), and observed to be free of heterozygous intervals. Chromosome arms were further defined as "focal" (the genomic regions analyzed here, namely the euchromatic portions of X, 2L, 2R, 3L, and 3R) or "non-focal" (the mitochondria and heterochromatin, including chromosomes 4 and Y). The assemblies analyzed here were defined as "release 2" data and are available for download at http://www.dpgp.org/dpgp2/DPGP2.html. Assemblies of mitochondrial and bacterial symbiont genomes are reported and analyzed separately [47].

*Estimation of consensus error rate*

Although the above assemblies provide nominal quality scores, we performed a separate evaluation of statistical confidence in the accuracy of assemblies. This analysis utilized five haploid embryo, reference strain ($y^1\ cn^1\ bw^1\ sp^1$) genomes resequenced with comparable depth and read characteristics as the non-reference genomes reported here (Table S2). In order to simulate the effects of genetic variation, the maq fakemut program [82] was used to introduce artificial substitutions and indels into the resequenced reference genomes.  Substitutions were introduced at rate 0.012/bp, while 1 bp indels were introduced at rate 0.0024/bp.  Alignment and consensus sequence generation was then performed as described above.

The artificially mutagenized reference genomes allowed us to examine the tradeoff between minimizing error rates and maximizing genomic coverage.  Based on the joint pattern of these quantities for various nominal quality scores (Figure S1), we selected a nominal quality threshold of Q31 as the basis for downstream analyses.  The observed consensus sequence error rate for the nominal Q31 cutoff suggested was equivalent to an average Phred score of Q48 (roughly one error per 100 kb).

*Detection of identical-by-descent genomic regions*

Long tracts of identity-by-descent (IBD) between genomes may result from the sampling of related individuals.  Because such relatedness violates the assumptions of many population genetic models, we sought to identify and mask instances of IBD caused by relatedness.  Target chromosomes from all possible pairs of genomes were compared to search for long intervals of identity-by-descent (IBD) that may result from close relatedness.  Following Langley *et al.* (28), windows 500 kb in length were moved in 100 kb increments across the genome, and sequence identity was defined as less than 0.0005 pairwise differences per site.  A large number of pairwise intervals fit this criterion (Table S3).  Some chromosomal intervals, including centromeres and telomeres, had recurrent IBD in between-population comparisons (Table S4).  Cross-population IBD occurred at scales up to 4 Mb within these manually delimited "recurrent IBD regions", and its occurrence between different populations suggests that processes other than close relatedness are responsible.  Such intervals were not masked from the data. We identified clear instances of "relatedness IBD" between two genomes when within-population IBD exceeded the scale observed between populations:  when more than 5 Mb of summed genome-wide IBD tracts occurred outside recurrent IBD regions, or when tracts greater than 5 Mb overlapped recurrent IBD regions.  Only nine pairs of genomes met one or both of these criteria (Table S4), and two of these pairs were expected based on the common origin of isofemale and chromosome extraction lines (Table S2).  For these pairs, one of the two genomes was chosen for filtering, and all identified IBD intervals from this pairwise comparison were masked to 'N' for most subsequent analyses.

*Admixture detection method – overview*

Relevant for the inference of non-African admixture is a panel of eight primary core genomes from France (the "FR" sample). *D. melanogaster* populations from outside sub-Saharan Africa show reduced genetic diversity and are more closely related to each other than to sub-Saharan populations [22,26]. Hence, whether admixture came from Europe or elsewhere in the diaspora, FR should represent an adequate "reference population" for the source of non-African admixture. However, we lack an African population that is known to be free of admixture. And while a variety of statistical methods exist for the detection of admixture, options for detecting unidirectional admixture using a single reference population are more limited. We therefore developed a new method to detect admixture in this data set.

We constructed a windowed Hidden Markov Model (HMM) machine learning approach based on a given haplotype's average pairwise divergence from the non-African reference population ($D_{FR}$). The admixed state is based on comparisons of individual FR haplotypes to the remainder of the FR sample. The non-admixed state is based on comparisons of haplotypes from a provisional "African panel" to the FR sample. Here, 22 genomes from the Rwanda "RG" sample are used as the African panel. We allow for the possibility of admixture within the African panel as described below.

Formally, the emissions distribution for the non-admixed state was constructed as follows. For each window, each RG haplotype was evaluated for average pairwise divergence from the FR sample ($D_{RG,FR}$). Each of these values was rescaled in terms of standard deviations of $D_{RG,FR}$ from the window mean $D_{RG,FR}$. Standardized values were added to the emissions distribution in bins of 0.1 standard deviations, and these bins were ultimately rescaled to sum to 1. Hence, the emissions distribution reflects the genome-wide pattern of $D_{RG,FR}$, accounting for local patterns of diversity.

The emissions distribution for the admixed state was constructed similarly. For each window, each FR haplotype was evaluated for average pairwise divergence from the remainder of the FR sample ($D_{FR,FR}$). However, these $D_{FR,FR}$ values were still rescaled by the window mean and standard deviation of $D_{RG,FR}$.  An alternative version of the method in which the admixed state's emissions distribution was instead rescaled by the local mean and standard deviation of $D_{FR,FR}$ was slightly less accurate when applied to simulated data.

Given these genome-wide emissions distributions, we can examine $D_{RG,FR}$ for each African allele for each window, and obtain its likelihood if we are truly making an "Africa-Europe comparison" with this $D_{RG,FR}$ (non-admixed state) or if we are actually making a "Europe-Europe comparison" (admixed state).  These likelihoods form the input for the HMM process, which was performed using an implementation [83] of the forward-backward algorithm.  A minimum admixture likelihood of 0.005 was applied to HMM input, in order to reduce the influence of a single unusual window.  Admixed intervals were defined as windows with >50% posterior probability for the admixed state.  For the purpose of masking admixed genomic intervals for downstream analyses, one window on each side of admixed intervals was added (to account for uncertainty in the precise boundaries of admixture tracts).

*Admixture detection method – validation*

The admixture detection method was tested using simulated data containing known admixture tracts.  Population samples of sequences 10 Mb in length were simulated using MaCS [84], which can approximate coalescent genealogies across long stretches of recombining sequence.  Demographic parameters were based on a published model for autosomal loci [13, 23].  The command line used was "./macs04 63 10000000 -s 12345 -i 1 -h 1000 -t 0.0376 -r 0.171 -c 5 86.5 -I 2 27 36 0 -en 0 2 0.183 -en 0.0037281 2 0.000377 -en 0.00381 2 1 -ej 0.00381000001 2 1 -eN 0.0145 0.2", specifying simulations with present population mutation rate 0.0376 and population recombination rate 0.171, gene conversion parameters based on a weighted average of loci from Yin *et al.* [85], and historic tree retention parameter $h$ = 1000 [84].

The above simulations generate population samples that may resemble data from sub-Saharan and cosmopolitan populations of *D. melanogaster*, but they do not involve any admixture.  If admixture was specified with the command line, then without modifications to the simulation program, there would not be an output record of admixture tract locations.  Instead, extra "non-African" haplotypes were simulated (one for each African haplotype), and these "donor alleles" became the source for admixture tracts which were spliced into the African population's data after MaCS simulation was completed.

The locations and lengths of admixture tracts were determined by a separate simulation process. The forward simulation program developed by Pool and Nielsen [36] accounts for drift, recombination, and migration, recording intervals with migrant history. By using this program to simulate a region symmetric to the African MaCS data, we identified intervals that should contain admixture tracts after $g$ generations of admixture. These intervals were then spliced from the non-African donor alleles into African haplotypes from the MaCS simulated polymorphism data.

The simulated data with admixture was then analyzed using the admixture HMM method described above. In this case, windows of 10 kb were analyzed. Times since the onset of admixture ($g$) of 100, 1000, and 10000 generations were examined. Migration rates were specified to approximate a total admixture proportion of 10% (hence testing the robustness of the method to this level of admixture in the "African panel").

As indicated by representative simulation results shown in Figure S8, the admixture detection method was highly accurate for $g$ = 100 and $g$ = 1000, and moderately accurate for $g$ = 10000. Based on preliminary observations from the data, we suspected that much of the admixture in our data set was on the order of $g$ = 100 or less.

*Admixture detection method – implementation*

The admixture HMM was initially applied to the RG sample alone. Compared with the simulated data, the empirical data showed more overlap between the admixed and non-admixed emissions distributions. This contrast could result from demographic differences between the African population used here (from Rwanda) and the one from which demographic parameter estimates were obtained (from Zimbabwe), and/or an effect of positive selection making Africa-Europe diversity comparisons more locally heterogeneous than expected under neutrality. We responded by expanding the window size used in the empirical data analysis. Windows were based on numbers of non-singleton polymorphic sites among the 22 RG primary core genomes. We chose a window size of 1000 such SNPs, which corresponds to a median window size close to 50 kb. Smaller windows led to noisier likelihoods (results not shown), while larger windows might exclude short admixture tracts without an appreciable gain in accuracy.

Another concern regarding the empirical data was the effect of sequencing depth on pairwise divergence values. After restricting the admixture analysis to genomes with >25X mean depth, we still observed a minor degree of "wavering" in admixture probabilities for genomes with the lowest depth. We therefore applied a simple correction factor to approximate each genome's quality effects on divergence metrics. In theory, we wish to know the effect of depth and other aspects of quality on $D_{FR}$. In practice, however, genomes differ in $D_{FR}$ in part based on their level of admixture. Instead, $D_{RG}$ (average pairwise divergence from the rest of the Rwanda sample) was used as a proxy. For each chromosome arm, a genome's $D_{RG}$ was compared to the RG population average. Each genome's $D_{FR}$ was then multiplied by the correction factor $\dfrac{\overline{D_{RG}}}{D_{RG}}$. Following this correction, no effect of depth on admixture inferences was observed within the primary core data set.

Although simulations suggested that our admixture method is robust to ~10% admixture in the African panel, we sought to maximize the method's accuracy by applying it iteratively to the RG sample. Identical-by-descent regions (as defined above) were masked during the creation of emissions distributions, but likelihoods were then evaluated for full RG chromosome arms. After one full "round" of the method (emissions, likelihoods, and HMM), admixture tracts were masked from the RG sample. This masked RG sample became the revised African panel for a second round of analysis, this one with a more accurate emissions distribution for the non-admixed state (since it contains more true "Africa-Europe" comparisons, and is presumably less influenced by admixture). Admixture masking for RG was redone based on round 2 admixture intervals, and the re-masked RG data was used to create a third and final set of emissions distributions. The round 3 emissions distributions were used to generate final admixture calls not only for the RG sample, but also for the other African genomes in the primary core data set.

The use of RG as an "African panel" when examining admixture in other African populations is not without concern. Fortunately, in addition to being the largest African sample, RG also occupies a genetically intermediate position within Africa (see results section), which reduces the potential impact of genetic structure on the accuracy of admixture inferences for non-RG genomes. It also appears that aside from the effects of admixture, no other African sample has a much closer relationship to FR than RG does (see results section), thus mitigating a potential source of bias.

*Analysis of admixture detection results*

Standard linear regression was used to investigate the possible relationship between cosmopolitan admixture proportion (for a population sample) and the human population size of the collection locality (city, town, or village population size). Census-based population estimates were obtained from online sources for 15 of 20 population samples. For the remainder, satellite-based estimates were obtained from fallingrain.com (Table S1). While a set of uniform and perfectly accurate population figures is not available for these locations, the estimates used here may still allow a significant effect of human population size on cosmopolitan admixture proportion to be detected.

The centiMorgan length of each admixture interval was calculated based on recombination rates inferred from smoothed genetic map data [28]. The extra buffer windows added to each side of conservative admixture tract delimitations described above were not included in these length estimates. CentiMorgan tract lengths were then used with a method [36] that estimates three parameters of a migration rate change model: the current migration rate, the previous migration rate, and the time of migration rate change. A minimum detectable tract length of 0.5 cM was chosen, corresponding to roughly 200 kb or 4 windows on average. Forward simulations [36] including recombination, migration, and drift were performed under the estimated demographic model. Simulated data were compared to empirical data, to test how often simulated variance in cosmopolitan admixture proportion exceeded that observed in the RG sample.

*Genetic diversity and structure of populations*

Regions of lower recombination proximal to centromeres and telomeres were excluded from most analyses, except where indicated below. Recombination rates were taken from mapping-based estimates [28], and the threshold between "low" and "high" recombination rates was defined as $2 \times 10^{-8}$ cross-overs per bp per generation. In most cases, a single transition point was apparent where a chromosome arm transitioned from low to high recombination, moving away from a centromere or telomere. A few narrow "valleys" of recombination rate estimates slightly below this threshold within broader high recombination regions, along with one peak of recombination rate slightly above this threshold close to the 3L centromere, were ignored in the definition of centromere-proximal and telomere-proximal boundaries. "Mid-chromosomal intervals" reflecting the higher recombination intervals used in this analysis for each chromosome arm are listed in Table SX.

Principal components analysis (PCA) was conducted using the method of Patterson *et al.* [43]. Mid-chromosomal data from all primary core genomes were included. The analysis was run twice, on data sets with and without admixture filtering. Applying additional filters (excluding sites with >5% missing data or <2.5% minor allele frequency) had little effect on results.

Nucleotide diversity ($\pi$) was initially calculated in 100 kb windows, and weighted values for each population sample (based on the number of sites in each window with data from at least two genomes) were then averaged to obtain a population's mean absolute $\pi$ for each chromosome arm. Relative $\pi$ was calculated by obtaining the ratio of window $\pi$ from a given population versus that for the RG population (the largest African sample), and window ratios were weighted by the number of sites with data from two or more RG genomes. Relative $\pi$ values should therefore be robust to cases where a population has large blocks of masked data in a genomic region with especially high or low diversity (since $\pi$ in each window is standardized by that observed for the RG sample), which could bias estimates of absolute $\pi$. Genome-wide relative $\pi$ was calculated as the unweighted average value of the five major chromosome arms. Three samples (CK, RC, SP) had only one primary core genome, but one or more secondary core genomes. Relative $\pi$ for these samples was calculated based on comparisons between primary and secondary core genomes, both for the target sample and for RG (which also contains primary and secondary core genomes). A similar re-estimation of relative $\pi$ for the CO sample yielded genome-wide relative $\pi$ of 0.914 from primary-secondary comparisons, versus 0.927 from primary core genomes only.

$D_{xy}$, the average rate of nucleotide differences between populations, was calculated for a subset of populations with high levels of genomic coverage in the admixture-filtered data (CO, ED, FR, GA, GU, KR, NG, RG, TZ, UG, ZI, ZS). $F_{ST}$ was calculated using the method of Hudson *et al.* [58], with equal population weightings regardless of their sample sizes. Arm-wide and genome-wide estimates of both statistics were calculated as described above for relative $\pi$.

Using the above summary statistics, we calculated the ratio of a population's $D_{ZI}$ (genetic distance from the four Zambia ZI genomes) to $\pi_{ZI}$. Here, the intention was to test which populations contained unique genetic diversity not observed in the maximally diverse ZI population, leading to ratios greater than one. The significance of ratios greater than one was assessed via a bootstrapping approach. Windows 100 kb in length were sampled with replacement until 667 were drawn, to match the number present in non-centromeric, non-telomeric regions of the empirical data. One million such replicates were conducted for each population, and the proportion of replicates with a ratio less than one became the bootstrapping *P* value. The use of windows much larger than the scale of linkage disequilibrium implies a conservative test.

For each population's genome-wide relative $\pi$ (Figure 6), and for the $D_{ZI}$ to $\pi_{ZI}$ ratio (Table 2; described), we applied a correction factor to reduce the predicted influence of sequencing depth on these quantities. From a linear regression of primary core genomes' sequencing depth *versus* $D_{ZI}$ (Figure 2), the slope and *y* intercept of this relationship were obtained. Based on population mean sequencing depth, a population's predicted $D_{ZI}$ was compared to the predicted $D_{ZI}$ of the reference population (RG for $\pi$, ZI for the ratio analysis). Observed summary statistic were multiplied by the ratio of these predicted values to obtain a corrected estimate. For both statistics, this adjustment led to changes of ~1% or less.

*Linkage disequilibrium from empirical and simulated data*

In addition to the standard correlation coefficient ($r^2$) of linkage disequilibrium (LD), we also examined directional LD via the $r_\omega$ statistic [70,71]. Here, LD is defined as positive if minor alleles preferentially occur on the same haplotype, and otherwise LD is negative. Empirical LD patterns were compared to data simulated under neutral evolution and equilibrium demography using *ms* [86]. In these simulations, the population mutation rate was taken from observed $\pi$. The population recombination rate was then inferred from the ratio of empirical estimates of recombination rates (the average rate from Langley *et al.* [28] for the analyzed X-linked and autosomal regions, simulated separately) and mutation rate [87]. Estimates for the rate of gene conversion relative to crossover events (5x) and the average gene conversion tract length (86.5 bp) were taken from a weighted average of the locus-specific estimates obtained by Yin *et al.* [85].

*Genomic scans for loci with unusual allele frequencies*

The $\Lambda_{max}$ statistic of Sweepfinder [75] uses allele frequencies to evaluate the relative likelihood of a selective sweep *versus* neutral evolution. To add information regard diversity reductions, we implemented the approach of Pavlidis *et al.* [76] to include a fraction of the invariant sites. One invariant site was added to the input for every 10 invariant sites that had <50% missing data. Likelihoods were evaluated for 1000 positions from each window. The folded allele frequency spectrum from short intron sites (see below) was used for background allele frequencies, assumed by the method to represent neutral evolution.

Local outliers for $\Lambda_{max}$ and $F_{ST}$ were examined in overlapping windows of 100 RG non-singleton SNPs (roughly 5 kb on average). For $F_{ST}$, overlapping windows were offset by increments of 20 RG non-singleton SNPs, in order to identify outlier loci that could result from adaptive population differentiation. Outlier windows were defined by the upper 2.5% ($F_{ST}$) or 5% ($\Lambda_{max}$) quantile for each chromosome arm. The lower threshold for $F_{ST}$ avoids an excessive number of outliers due to the greater number of (overlapping) windows, compared to the non-overlapping windows for $\Lambda_{max}$. Outliers with up to two non-overlapping non-outlier windows between them were considered as part of the same "outlier region", since they might reflect a single evolutionary signal. For $F_{ST}$, the center of an outlier region was defined as the midpoint of its most extreme window. The nearest gene to an outlier region was calculated based on the closest exon (protein-coding or untranslated) to the above location, based on *D. melanogaster* genome release 5.43 coordinates obtained from Flybase.

Two $F_{ST}$ outlier analyses were conducted. One, with the aim of identifying loci that may have contributed to the adaptive difference between African and cosmopolitan populations, focused on $F_{ST}$ between the FR and RG population samples. The other scan was intended to search for potential adaptive differences among African populations. The nine population samples with a mean post-filtering sample size above 3.75 were included (CO, ED, GA, GU, NG, RG, UG, ZI, ZS). The mean $F_{ST}$ from all pairwise population comparisons was evaluated for each window, and outlier regions for this overall $F_{ST}$ were obtained. Each population was also analyzed separately, in terms of the mean $F_{ST}$ from eight pairwise population comparisons. Here, outliers were analyzed separately for each African population, but the lists of population-specific outliers were also combined for more statistically powerful enrichment tests.

The enrichment of gene ontology (GO) categories among sets of outliers was evaluated. For each GO category, the number of unique genes that were the closest to an outlier region center (see above) was noted. A *P* value was then calculated, representing the probability of observing as many (or more) outlier genes from that category under the null hypothesis of a random distribution of outlier region centers across all windows. Calculating null probabilities based on windows, rather than treating each gene identically, accounts for the fact that genes vary greatly in length, and hence in the number of windows that they are associated with. *P* values were obtained from a permutation approach in which all outlier region center windows were randomly reassigned 10,000 times (results not shown).

LITERATURE CITED

1. Adams MD, Celniker SE, Holt RA, Evans CA, Gocayne JD, *et al.* (2000) The genome sequence of *Drosophila melanogaster*. Science 287: 2185-2195.

2. Roy S, Ernst J, Kharchenko PV, Kheradpour P, Negre N, *et al.* (2010) Identification of functional elements and regulatory circuits by *Drosophila* modENCODE. Science 330: 1787-1797.

3. Dobzhansky T, Sturtevant AH (1938) Inversions in the Chromosomes of *Drosophila pseudoobscura*. Genetics 23: 28-64.

4. Lewontin RC, Hubby JL (1966) A molecular approach to the study of genic heterozygosity in natural populations. II. Amount of variation and degree of heterozygosity in natural populations of *Drosophila pseudoobscura*. Genetics 54: 595-609.

5. Singh RS, Hickey DA, David J (1982) Genetic differentiation between geographically distant populations of *Drosophila melanogaster*. Genetics 101: 235-256.

6. Mettler LE, Voelker RA, Mukai T (1977) Inversion clines in populations of *Drosophila melanogaster*. Genetics 87: 169-176

7. Hudson RR, Kreitman M, Aguadé (1987) A test of neutral molecular evolution based on nucleotide data. Genetics 116: 153-159.

8. McDonald JH, Kreitman M (1991) Adaptive protein evolution at the Adh locus in *Drosophila*. Nature 351: 652–654.

9. Begun DJ, Aquadro CF (1992) Levels of naturally occurring DNA polymorphism correlate with recombination rates in *D. melanogaster*. Nature 356: 519–520.

10. Maynard Smith J, Haigh J (1974) The hitch-hiking effect of a favourable gene. Genet Res 23: 23–35.

11. Charlesworth B, Morgan MT, Charlesworth D (1993) The effect of deleterious mutations on neutral molecular variation. Genetics 134: 1289–1303.

12. Kauer M. O., Dieringer D, Schlötterer C (2003) A microsatellite variability screen for positive selection associated with the "out of Africa" habitat expansion of *Drosophila melanogaster*.

13. Hutter S, Li H, Beisswanger S, De Lorenzo D, Stephan W (2007) Distinctly different sex ratios in African and European populations of *Drosophila melanogaster* inferred from chromosomewide single nucleotide polymorphism data. Genetics 177: 469-480.

14. Smith NGC, Eyre-Walker A (2002) Adaptive protein evolution in Drosophila. Nature 415: 1022–1024.

15. Andolfatto P (2005) Adaptive evolution of non-coding DNA in *Drosophila*. Nature 437: 1149–1152.

16. Lachaise D, Cariou ML, David JR, Lemeunier F, Tsacas L, et al. (1988) Historical biogeography of the *Drosophila melanogaster* species subgroup. In: Hecht MK, Wallace B, Prance GT, eds. Evolutionary biology. New York: Plenum. pp. 159–225.

17. Begun DJ, Aquadro CF (1993) African and North American populations of *Drosophila melanogaster* are very different at the DNA level. Nature 365: 548–550.

18. Pool JE, Aquadro CF (2006) History and structure of sub-Saharan populations of *Drosophila melanogaster*. Genetics 174: 915-929.

19. Lachaise D, Silvain J-F (2004) How two Afrotropical endemics made two cosmopolitan commensals: the *Drosophila melanogaster* – *D. simulans* paleogeographic riddle. Genetica 120: 17-39.

20. Veuille M, Baudry E, Cobb M, Derome N, Gravot E (2004) Historicity and the population genetics of *Drosophila melanogaster* and *D. simulans*. Genetica 120: 61-70.

21. Haddrill PR, Thornton KR, Charlesworth B, Andolfatto P (2005) Multilocus patterns of nucleotide variability and the demographic and selection history of *Drosophila* melanogaster populations. Genome Res 15: 790–799.

22. Baudry E, Viginier B, Veuille M (2004) Non-African populations of *Drosophila melanogaster* have a unique origin. Mol Biol Evol 21: 1482-1491.

23. Li H, Stephan W (2006) Inferring the demographic history and rate of adaptive substitution in *Drosophila*. PLoS Genet 2: e166.

24. Thornton KR, Andolfatto P (2006) Approximate Bayesian inference reveals evidence for a recent, severe bottleneck in a Netherlands population of *Drosophila melanogaster*. Genetics 172: 1607–1619.

25. Lintner JA (1882) First annual report on the injurious and other insects of the State of New York. Albany, New York: Weed, Parsons, and Co.

26. Caracristi G, Schlötterer C (2003) Genetic differentiation between American and European *Drosophila melanogaster* populations could be attributed to admixture of African alleles. Mol Biol Evol 20: 792-799.

27. Begun DJ, Holloway AK, Stevens K, Hillier LW, Poh YP, et al. (2007) Population genomics: whole-genome analysis of polymorphism and divergence in *Drosophila simulans*. PLoS Biol 5: e310.

28. Langley CH, Stevens K, Cardeno C, Lee YCG, Schrider DR, *et al.* (2012) Genomic variation in natural populations of *Drosophila melanogaster*. Genetics, In Press.

29. Mackay TFC, Richards S, Stone EA, Barbadilla A, Ayroles JF, *et al.* (2012) The *Drosophila melanogaster* genetic reference panel. Nature 482: 173-178.

30. Langley CH, Crepeau M, Cardeno C, Corbett-Detig R, Stevens K (2011) Circumventing heterozygosity: sequencing the amplified genome of a single haploid *Drosophila melanogaster* embryo. Genetics 188: 239-246.

31. Li H, Durbin R (2009) Fast and accurate short read alignment with Burrows-Wheeler transform. Bioinformatics 25: 2078-2079.

32. Li H, Handsaker B, Wysoker A, Fennell T, Ruan J, *et al.* (2009) The Sequence Alignment / Map format and SAMtools. Bioinformatics 25: 2078-2079.

33. Capy P, Veuille M, Paillette M, Jallon J-M, Vouidibio J, *et al.* (2000) Sexual isolation of genetically differentiated sympatric populations of *Drosophila melanogaster* in Brazzaville, Congo: the first step towards speciation? Heredity 84: 468-475.

34. Kauer M, Dieringer D, Schlötterer C (2003) Nonneutral admixture of immigrant genotypes in African *Drosophila melanogaster* populations from Zimbabwe. Mol. Biol. Evol. 20:1329-1337.

35. Vouidibio J, Capy P, Defaye D, Pla E, Sandrin E, *et al.* (1989) Short-range genetic structure of *Drosophila melanogaster* populations in an Afrotropical urban area and its significance. Proc Natl Acad Sci USA 86: 8442-8446.

36. Pool JE, Nielsen R (2009) Inference of historical changes in migration rate from the lengths of migrant tracts. Genetics 181: 711-719.

37. Catania F, Kauer MO, Daborn PJ, Yen JL, Ffrench-Constant RH, *et al.* (2004) A world-wide survey of an *Accord* insertion and its association with DDT resistance in *Drosophila melanogaster*. Mol Ecol 13: 2491-2504.

38. Wu C-I, Hollocher H, Begun DJ, Aquadro CF, Xu Y, *et al.* (1995) Sexual isolation in *Drosophila melanogaster*: a possible case of incipient speciation. Proc Natl Acad Sci USA 92: 2519-2523.

39. Hollocher H, Ting C-T, Pollack F, Wu C-I (1997) Incipient speciation by sexual isolation in *Drosophila melanogaster*: variation in mating preference and correlation between the sexes. Evolution 51: 1175-1181.

40. Charlesworth B, Coyne JA, Barton NH (1987) The relative rates of evolution of sex chromosomes and autosomes. Am Nat 130: 113-146.

41. Aulard S, David JR, Lemeunier F (2002) Chromosomal inversion polymorphism in Afrotropical populations of *Drosophila melanogaster*. Genet Res 79: 49-63.

42. Lucchesi JC, Suzuki DT (1968) The interchromosomal control of recombination. Ann Rev Genet 2: 53-86.

43. Patterson N, Price AL, Reich D (2006) Population structure and eigenanalysis. PLoS Genet 2: e190.

44. Sackton TB, Kulathinal RJ, Bergman CM, Quinlan AR, Dopman EB, *et al.* (2009) Population genomic inferences from sparse high-throughput sequencing of two populations of *Drosophila melanogaster*. Genome Biol Evol 1: 449-465.

45. van Herrewege J, David JR (1997) Starvation and desiccation tolerances in *Drosophila*: comparison of species from different climatic origins. Ecoscience 4: 151-157.

46. Stanley SM, Parsons PA, Spence GE, Weber L (1980) Resistance of species of the *Drosophila melanogaster* subgroup to environmental extremes. Aust J Zool 28: 413-421.

47. Richardson MF, Weinert LM, Welch JJ, Linheiro RS, Magwire MM, *et al.* (2012) Population genomics of the *Wolbachia* endosymbiont in *Drosophila melanogaster*. (Joint Submission)

48. Corbett-Detig RB, Hartl DL (2012) Population genomics of inversion polymorphisms in *Drosophila melanogaster*. (Joint Submission)

49. Guerrero RF, Rousset F, Kirkpatrick M (2012) Coalescent patterns for chromosomal inversions in divergent populations. *Phil Trans R Soc B* 367:430-438.

50. Hasson E, Eanes WF (1996) Contrasting histories of three gene regions associated with *In(3L)Payne* of *Drosophila melanogaster*. Genetics 144:1565-1575.

51. Andolfatto P, Przeworski M (2001) Regions of lower crossing over harbor more rare variants in African populations of Drosophila melanogaster. Genetics 158: 657–665.

52. Greenberg R, Crow JF (1960) A comparison of the effect of lethal and detrimental chromosomes from *Drosophila* populations. Genetics 8:1153-1168.

53. Dobzhansky T (1954) Evolution as a creative process. Proceedings of the 9th International Congress on Genetics, Bellagio, Italy, 1:435-449.

54. Ohta T (1971) Associative overdominance caused by linked detrimental mutations. Genet Res 18:277-286.

55. Kirkpatrick M, Barton N (2006) Chromosome inversions, local adaptation and speciation. Genetics 173:419-434.

56. Bierne N, Tsitrone A, David P (2000) An inbreeding model of associative overdominance during a population bottleneck. Genetics 155:1981-1990.

57. Pool JE, Nielsen R (2008) The impact of founder events on chromosomal variability in multiply mating species. Mol Biol Evol 25: 1728-1736.

58. Hudson RR, Slatkin M, Maddison WP (1992) Estimation of levels of gene flow from DNA sequence data. Genetics 132: 583-589.

59. Halligan DL, Keightley PD (2006) Ubiquitous selective constraints in the *Drosophila* genome revealed by a genome-wide interspecies comparison. Genome Res 16: 875–884.

60. Parsch J, Novozhilov S, Saminadin-Peter SS, Wong KM, Andolfatto P. On the utility of short intron sequences as a reference for the detection of positive and negative selectioni in *Drosophila*. (2010) Mol Biol Evol 27:1226-1234,

61. Sella G, Petrov DA, Przeworski M, Andolfatto P (2009) Pervasive natural selection in the *Drosophila* genome? PLoS Genet 5: e1000495.

62. Jensen JD, Thornton KR, Andolfatto P (2008) Approximate Bayesian estimator suggests strong, recurrent selective sweeps in *Drosophila*. PLoS Genet 4: e1000198.

63. Pool JE, Nielsen R (2007) Population size changes reshape genomic patterns of diversity. Evolution 61: 3001-3006.

64. Braverman JM, Hudson RR, Kaplan NL, Langley CH, Stephan W (1995) The hitchhiking effect on the site frequency spectrum of DNA polymorphisms. Genetics 140: 783–796.

65. Pennings PS, Hermisson J (2006) Soft sweeps III: the signature of positive selection from recurrent mutation. PLoS Genet 2: e186.

66. Charlesworth D, Charlesworth B, Morgan MT (1995) The pattern of neutral molecular variation under the background selection model. Genetics 141: 1619–1632.

67. Zeng K, Charlesworth B (2011) The joint effects of background selection and genetic recombination on local gene genealogies. Genetics 189:251-266.

68. Ometto L, Glinka S, De Lorenzo D, Stephan W (2005) Inferring the effects of demography and selection on *Drosophila melanogaster* populations from a chromosome-wide scan of DNA variation. Mol Biol Evol 22: 2119–2130.

69. Shapiro JA, Huang W, Zhang C, Hubisz MJ, Lu J, *et al.* (2007) Adaptive genic evolution in the *Drosophila* genomes. Proc Natl Acad Sci USA 104: 2271-2276.

70. Langley CH, Crow JF (1974) The direction of linkage disequilibrium. Genetics 78: 937-941.

71. Langley CH, Tobari YN, Kojima K-I (1974) Linkage disequilibrium in natural populations of *Drosophila melanogaster*. Genetics 78: 921-936.

72. Stephan W, Song YS, Langley CH (2006) The hitchhiking effect on linkage disequilibrium between linked neutral loci. Genetics 172: 2647-2663.

73. Chan AH, Jenkins P, Song YS (2012) Genome-wide fine-scale recombination rate variation in *Drosophila melanogaster*. (Joint Submission)

74. Jensen JD, Kim Y, Bauer DuMont V, Aquadro CF, Bustamante CD (2005) Distinguishing between selective sweeps and demography using DNA polymorphism data. Genetics 170: 1401–1410.

75. Nielsen R, Williamson S, Kim Y, Hubisz MJ, Clark AG, et al. (2005) Genomic scans for selective sweeps using SNP data. Genome Res 15: 1566–1575.

76. Pavlidis P, Jensen JD, Stephan W (2010) Searching for footprints of positive selection in whole-genome SNP data from nonequilibrium populations. Genetics 185: 907-922.

77. Aguadé M (2009) Nucleotide and copy-number polymorphism at the odorant receptor genes *Or22a* and *Or22b* in *Drosophila melanogaster*. Mol Biol Evol 26: 61-70.

78. Aminetzach YT, Macpherson JM, Petrov DA (2005) Pesticide resistance via transposition-mediated adaptive gene truncation in Drosophila. Science 309: 764–767.

79. Magwire MM, Bayer F, Webster CL, Cao C, Jiggins FM. (2011) Successive increases in the resistance of *Drosophila* to viral infection through a transposon insertion followed by a duplication. PLoS Genet 7:e1002337.

80. Pool JE (2009) Notes regarding the collection of African *Drosophila melanogaster*. Dros Inf Serv 92:130-134.

81. Fuyama Y (1984) Gynogenesis in *Drosophila melanogaster*. Jpn J Genet 59: 91-96.

82. Li H, Ruan J, Durbin R (2008) Mapping short DNA sequencing reads and calling variants using mapping quality scores. Genome Res 18: 1851-1858.

83. Emerson JJ, Cardoso-Moreira M, Borevitz JO, Long M (2008) Natural selection shapes genome-wide patterns of copy-number polymorphism in *Drosophila melanogaster*. Science 320: 1629-1631.

84. Chen GK, Marjoram P, Wall JD (2009) Fast and flexible simulation of DNA sequence data. Genome Res 19: 136-142.

85. Yin J, Jordan MI, Song YS (2009) Joint estimation of gene conversion rates and mean conversion tract lengths from population SNP data. Bioinformatics 25: i231-i239.

86. Hudson RR (2002) Generating samples under a Wright-Fisher neutral model. Bioinformatics 18: 337-338.

87. Keightley PD, Trivedi U, Thomson M, Oliver F, Kumar S, *et al.* (2009) Analysis of the genome sequences of three *Drosophila melanogaster* spontaneous mutation accumulation lines. Genome Res 19: 1195-1201.

NOTE:

Supplemental figures and tables are not included with this submission, but can be downloaded at: http://www.johnpool.net/Pool2012supplemental.zip

| Population | X | 2L | 2R | 3L | 3R | Average |
|---|---|---|---|---|---|---|
| CK* | 0.77 | ** | ** | 0.92 | 0.93 | 0.87 |
| CO | 0.88 | 0.97 | 0.94 | 0.97 | 0.87 | 0.93 |
| ED | 0.73 | 0.83 | 0.82 | 0.86 | 0.77 | 0.80 |
| EZ | 0.78 | 0.85 | 0.83 | 0.80 | 0.90 | 0.83 |
| FR | 0.41 | 0.66 (0.58) | 0.59 | 0.75 (0.62) | 0.84 (0.63) | 0.65 |
| GA | 0.94 | 0.98 | 1.01 | 1.05 | 1.07 | 1.01 |
| GU | 0.91 | 1.00 | 0.98 | 1.00 | 1.02 | 0.98 |
| KN | 1.00 | 0.78 | 1.00 | 1.02 | 1.11 | 0.98 |
| KR | 0.96 | 0.72 | 1.03 | 0.99 | 1.08 | 0.96 |
| KT | 1.00 | 1.05 | 0.99 | 1.03 | 1.03 | 1.02 |
| NG | 0.91 | 0.90 | 0.93 | 0.96 | 1.00 | 0.94 |
| RC* | 1.04 | 0.98 | 0.98 | 1.00 | 0.98 | 0.99 |
| SP* | 1.05 | 1.08 | 1.05 | 1.01 | 0.86 | 1.01 |
| TZ | 0.66 | 0.68 | 1.02 | 0.93 | 1.01 | 0.86 |
| UG | 0.98 | 1.02 | 1.00 | 1.02 | 0.94 | 0.99 |
| UM | 0.96 | 0.99 | 1.05 | 1.02 | 1.07 | 1.02 |
| ZI | 1.05 | 1.15 | 1.07 | 1.10 | 1.17 | 1.11 |
| ZO | 1.05 | ** | 1.03 | 1.03 | 1.06 | 1.04 |
| ZS | 0.96 | 1.13 | 1.03 | 1.00 | 1.13 | 1.05 |

**Table 1.**  Relative nucleotide diversity (versus the RG sample) for each population sample is given for chromosome arms and the average of arms.  Data consisted of non-centromeric, non-telomeric regions, with putatively admixed regions masked from African genomes.  For the FR sample, values in parentheses reflect the exclusion of inverted chromosomes.
* denotes a value based on comparisons between primary and secondary core genomes.
** indicates arms for which diversity could not be estimated due to a lack of non-masked data.

| Population | CO | ED | FR | GA | GU | KR | NG | RG | TZ | UG | ZI | ZS |
|---|---|---|---|---|---|---|---|---|---|---|---|---|
| CO | **0.702** | 0.780 | 0.765 | 0.759 | 0.745 | 0.759 | 0.738 | 0.770 | 0.781 | 0.766 | 0.841 | 0.811 |
| ED | 0.159 | **0.614** | 0.781 | 0.801 | 0.790 | 0.778 | 0.783 | 0.789 | 0.799 | 0.786 | 0.845 | 0.822 |
| FR | 0.224 | 0.297 | **0.491** | 0.774 | 0.772 | 0.751 | 0.764 | 0.783 | 0.783 | 0.779 | 0.828 | 0.805 |
| GA | 0.035 | 0.143 | 0.193 | **0.763** | 0.768 | 0.770 | 0.749 | 0.789 | 0.793 | 0.789 | 0.858 | 0.827 |
| GU | 0.031 | 0.144 | 0.205 | 0.020 | **0.741** | 0.769 | 0.743 | 0.781 | 0.790 | 0.778 | 0.851 | 0.821 |
| KR | 0.077 | 0.156 | 0.205 | 0.052 | 0.063 | **0.707** | 0.744 | 0.755 | 0.700 | 0.763 | 0.810 | 0.760 |
| NG | 0.048 | 0.161 | 0.221 | 0.020 | 0.028 | 0.055 | **0.703** | 0.772 | 0.772 | 0.772 | 0.843 | 0.809 |
| RG | 0.056 | 0.135 | 0.208 | 0.039 | 0.043 | 0.037 | 0.057 | **0.754** | 0.771 | 0.763 | 0.828 | 0.800 |
| TZ | 0.138 | 0.214 | 0.274 | 0.113 | 0.123 | 0.037 | 0.127 | 0.091 | **0.650** | 0.784 | 0.819 | 0.754 |
| UG | 0.052 | 0.135 | 0.206 | 0.041 | 0.042 | 0.047 | 0.058 | 0.015 | 0.110 | **0.750** | 0.838 | 0.811 |
| ZI | 0.090 | 0.146 | 0.205 | 0.072 | 0.077 | 0.053 | 0.090 | 0.043 | 0.094 | 0.057 | **0.831** | 0.817 |
| ZS | 0.082 | 0.149 | 0.208 | 0.063 | 0.070 | 0.015 | 0.078 | 0.036 | 0.046 | 0.052 | 0.008 | **0.790** |
| $D_{ZI}/\pi_{ZI}$: | 1.023* | 1.027* | 1.001 | 1.026* | 1.029* | 0.990 | 1.022* | 1.003 | 0.999 | 1.015* | (1) | 0.988 |

**Table 2.** Nucleotide diversity and genetic differentiation are shown, averaged across the non-centromeric, non-telomeric regions of each chromosome arm. Values above the diagonal represent $D_{xy}$ (in percent), while those below reflect $F_{ST}$. Bold values on the diagonal are $\pi$ (%). The ratio of each population's genetic distance to the ZI sample versus diversity with the ZI sample is also given (bottom row). Ratios were corrected based on the (minor) predicted effects of sequencing depth for each population (see Materials and Methods). Ratios significantly greater than one (bootstrapping $P < 0.001$) are noted (*). Admixture-filtered data from genomes with less than 15% estimated admixture were analyzed for each population that had two or more such genomes.

| Gene Ontology Category Description | Outlier Genes | Total Genes | P value |
|---|---|---|---|
| mRNA binding | 22 | 120 | 0 |
| microtubule associated complex | 20 | 184 | 0 |
| lipid particle | 15 | 148 | 0 |
| polytene chromosome | 9 | 37 | 0 |
| mRNA 3'-UTR binding | 7 | 13 | 0 |
| ribonucleoprotein complex | 6 | 12 | 0 |
| positive regulation of translation | 4 | 7 | 0 |
| heterochromatin | 4 | 8 | 0 |
| nuclear pore | 6 | 21 | 0.0001 |
| male meiosis | 5 | 18 | 0.0001 |
| SMAD protein import into nucleus | 4 | 10 | 0.0001 |
| ubiquitin-protein ligase activity | 8 | 48 | 0.0002 |
| precatalytic spliceosome | 8 | 73 | 0.0002 |
| nuclear mRNA splicing, via spliceosome | 9 | 102 | 0.0003 |
| nucleus | 65 | 699 | 0.0004 |
| polytene chromosome puff | 4 | 15 | 0.0005 |
| regulation of alternative nuclear mRNA splicing, via spliceosome | 7 | 34 | 0.0006 |
| neurogenesis | 22 | 316 | 0.0008 |
| female meiosis chromosome segregation | 5 | 21 | 0.0013 |
| positive regulation of transcription, DNA-dependent | 9 | 34 | 0.0014 |
| DNA binding | 25 | 248 | 0.0017 |
| salivary gland cell autophagic cell death | 8 | 39 | 0.0023 |
| protein ubiquitination | 4 | 15 | 0.0024 |
| nucleoplasm | 4 | 14 | 0.0028 |
| spermatogenesis | 8 | 56 | 0.0032 |
| mitotic cell cycle | 4 | 17 | 0.0036 |
| regulation of apoptotic process | 4 | 11 | 0.0039 |
| regulation of mitosis | 4 | 18 | 0.0049 |
| chromatin organization | 4 | 13 | 0.0051 |
| negative regulation of transcription, DNA-dependent | 9 | 41 | 0.0057 |
| cytokinesis | 7 | 42 | 0.0059 |
| phagocytosis, engulfment | 13 | 112 | 0.0068 |
| autophagic cell death | 7 | 36 | 0.0073 |
| protein complex | 7 | 36 | 0.0079 |
| fusome | 5 | 16 | 0.0083 |
| nuclear envelope | 4 | 18 | 0.0086 |

**Table 3.** Gene ontology enrichment analysis based on outlier windows for high $\Lambda_{max}$ in the Rwanda RG sample, indicating potential targets of recent selective sweeps. Listed are GO categories with $P < 0.01$ and outlier genes > 3. Full results are given in Table S14.

| Gene Ontology Category Description | Outlier Genes | Total Genes | P value |
|---|---|---|---|
| DNA-directed RNA polymerase activity | 3 | 17 | 0.00103 |
| oocyte microtubule cytoskeleton organization | 3 | 7 | 0.0033 |
| regulation of alternative nuclear mRNA splicing, via spliceosome | 6 | 33 | 0.00416 |
| olfactory receptor activity | 6 | 32 | 0.00419 |
| mitochondrial matrix | 4 | 31 | 0.00485 |
| positive regulation of protein phosphorylation | 2 | 5 | 0.00519 |
| regulation of adult chitin-containing cuticle pigmentation | 3 | 8 | 0.00638 |
| regulation of R8 cell spacing in compound eye | 3 | 4 | 0.00786 |
| notum cell fate specification | 3 | 3 | 0.00834 |
| receptor signaling protein serine/threonine kinase activity | 4 | 17 | 0.0096 |
| regulation of nuclear mRNA splicing, via spliceosome | 2 | 6 | 0.01232 |
| sensory perception of smell | 8 | 49 | 0.01485 |
| RNA polymerase II transcription cofactor activity | 2 | 12 | 0.01549 |
| mRNA binding | 13 | 114 | 0.01581 |
| mediator complex | 2 | 14 | 0.01648 |
| nucleobase-containing compound metabolic process | 2 | 6 | 0.01681 |
| SMAD protein import into nucleus | 2 | 10 | 0.01889 |
| transcription from RNA polymerase II promoter | 3 | 20 | 0.01932 |
| lipid particle | 8 | 138 | 0.02089 |
| muscle cell homeostasis | 3 | 7 | 0.02217 |
| spermatocyte division | 2 | 6 | 0.02634 |
| embryonic axis specification | 2 | 5 | 0.02869 |
| cytosolic small ribosomal subunit | 2 | 24 | 0.02915 |
| haltere development | 2 | 3 | 0.03257 |
| MAPK cascade | 2 | 8 | 0.03302 |
| mucosal immune response | 2 | 5 | 0.03376 |
| odorant binding | 6 | 61 | 0.03402 |
| dendrite | 3 | 19 | 0.03486 |
| small nuclear ribonucleoprotein complex | 2 | 22 | 0.03584 |
| notum development | 2 | 3 | 0.03783 |
| neurexin family protein binding | 2 | 2 | 0.03902 |
| induction of apoptosis | 2 | 9 | 0.0406 |
| myofibril assembly | 2 | 4 | 0.04232 |
| oocyte axis specification | 3 | 11 | 0.04339 |

**Table 4.** Gene ontology enrichment analysis based on outlier windows for high mean $F_{ST}$ for African population comparisons. Listed are GO categories with $P < 0.05$ and outlier genes > 1. Full results are given in Table S16.

| Gene Ontology Category Description | Outlier Genes | Total Genes | P value |
|---|---|---|---|
| chromosome segregation | 5 | 20 | 0.00106 |
| dephosphorylation | 3 | 11 | 0.00315 |
| digestion | 2 | 4 | 0.00389 |
| locomotion | 4 | 8 | 0.00601 |
| basal lamina | 3 | 5 | 0.00675 |
| polytene chromosome interband | 3 | 17 | 0.0087 |
| pyruvate metabolic process | 2 | 6 | 0.00879 |
| female germ-line cyst formation | 2 | 3 | 0.01018 |
| GTPase activity | 8 | 76 | 0.01732 |
| regulation of protein localization | 2 | 4 | 0.01821 |
| tissue development | 2 | 4 | 0.01851 |
| iron ion binding | 3 | 17 | 0.01888 |
| organ morphogenesis | 2 | 4 | 0.01895 |
| FMN binding | 2 | 7 | 0.02285 |
| actin filament bundle assembly | 3 | 8 | 0.02419 |
| histone phosphorylation | 2 | 5 | 0.02495 |
| nucleus localization | 2 | 3 | 0.0271 |
| germ cell development | 4 | 19 | 0.03221 |
| eye development | 3 | 6 | 0.03279 |
| ATPase activity, coupled | 6 | 40 | 0.03319 |
| alcohol metabolic process | 3 | 12 | 0.03355 |
| organic anion transport | 2 | 7 | 0.03944 |
| metal ion binding | 5 | 44 | 0.0395 |
| organic anion transmembrane transporter activity | 2 | 7 | 0.03967 |
| mitotic cell cycle | 3 | 17 | 0.04069 |
| transcription coactivator activity | 3 | 8 | 0.04122 |
| larval chitin-based cuticle development | 2 | 5 | 0.04311 |
| lipid particle | 8 | 138 | 0.04768 |
| anion transport | 2 | 4 | 0.04834 |
| neuropeptide receptor activity | 8 | 30 | 0.04931 |
| choline dehydrogenase activity | 3 | 13 | 0.04994 |

**Table 5.** Gene ontology enrichment analysis based on outlier windows for high $F_{ST}$ between Rwanda and France population samples. Listed are GO categories with $P < 0.05$ and outlier genes > 1. Full results are given in Table S18.

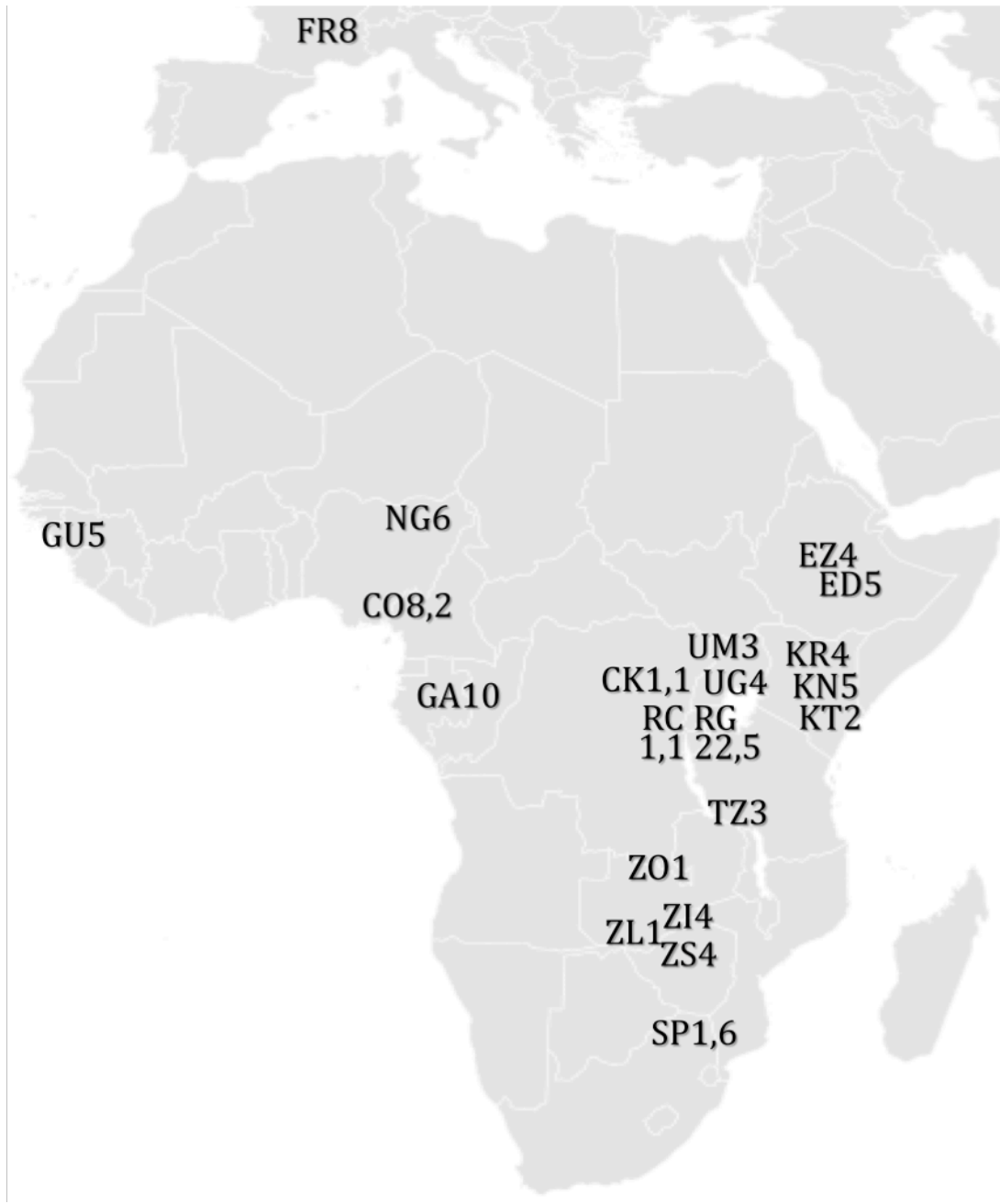

**Figure 1.** Locations of population samples from which the analyzed genomes were derived. Each population sample is indicated by a two letter abbreviation followed by the number of primary core genomes sequenced. For populations with secondary core genomes, that number follows a comma. Additional data and sample characteristics are described in Table S1.

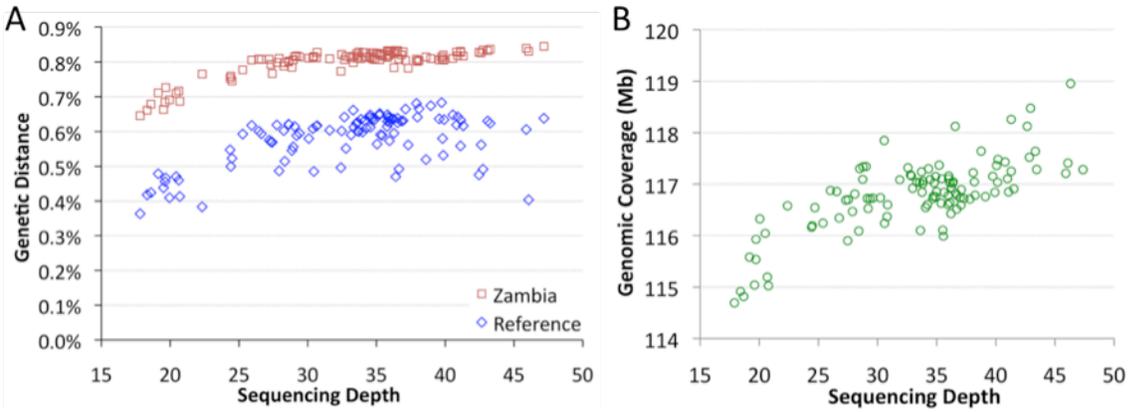

**Figure 2.** Mean sequencing depth is correlated with genetic distance (A) and genomic coverage (B). African core genomes with data from all major chromosome arms are depicted. The effect of depth on genetic distance applies whether genomes are compared to the published reference genome (blue) or the Zambia ZI population sample (red). Subsequent analyses focused largely on "primary core" genomes with >25X depth.

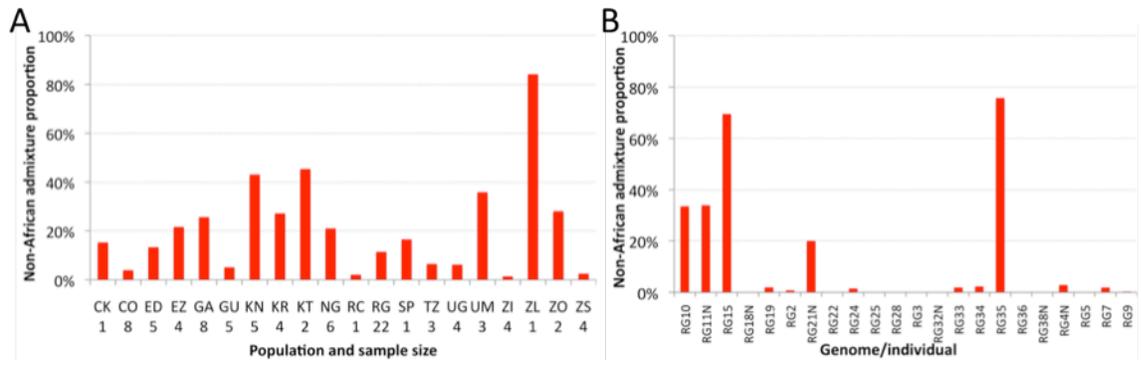

**Figure 3.** Heterogeneity in estimated cosmopolitan admixture proportions, both among African populations (A) and within the Rwanda RG population sample (B).

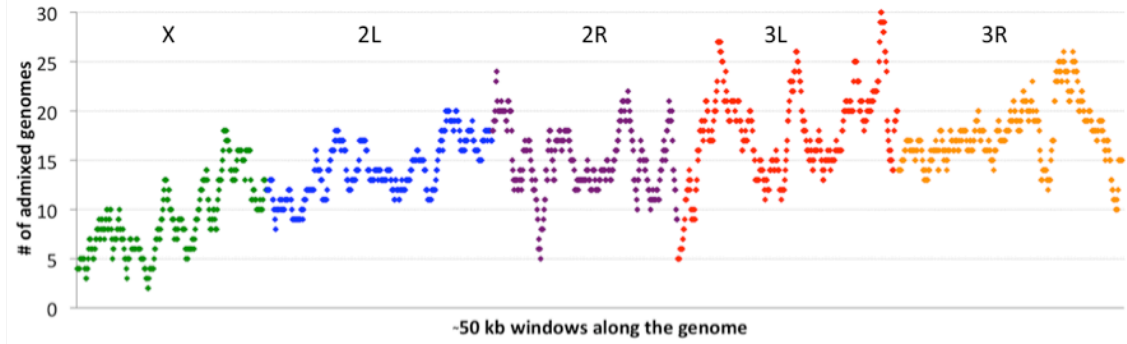

**Figure 4.** Cosmopolitan admixture levels are depicted across the genome. For each genomic window, the number of African primary core genomes (across all populations) with >50% admixture probability is plotted. Chromosome arms are labeled and indicated by color. Each window contains 1000 RG non-singleton SNPs (approximately 50 kb on average).

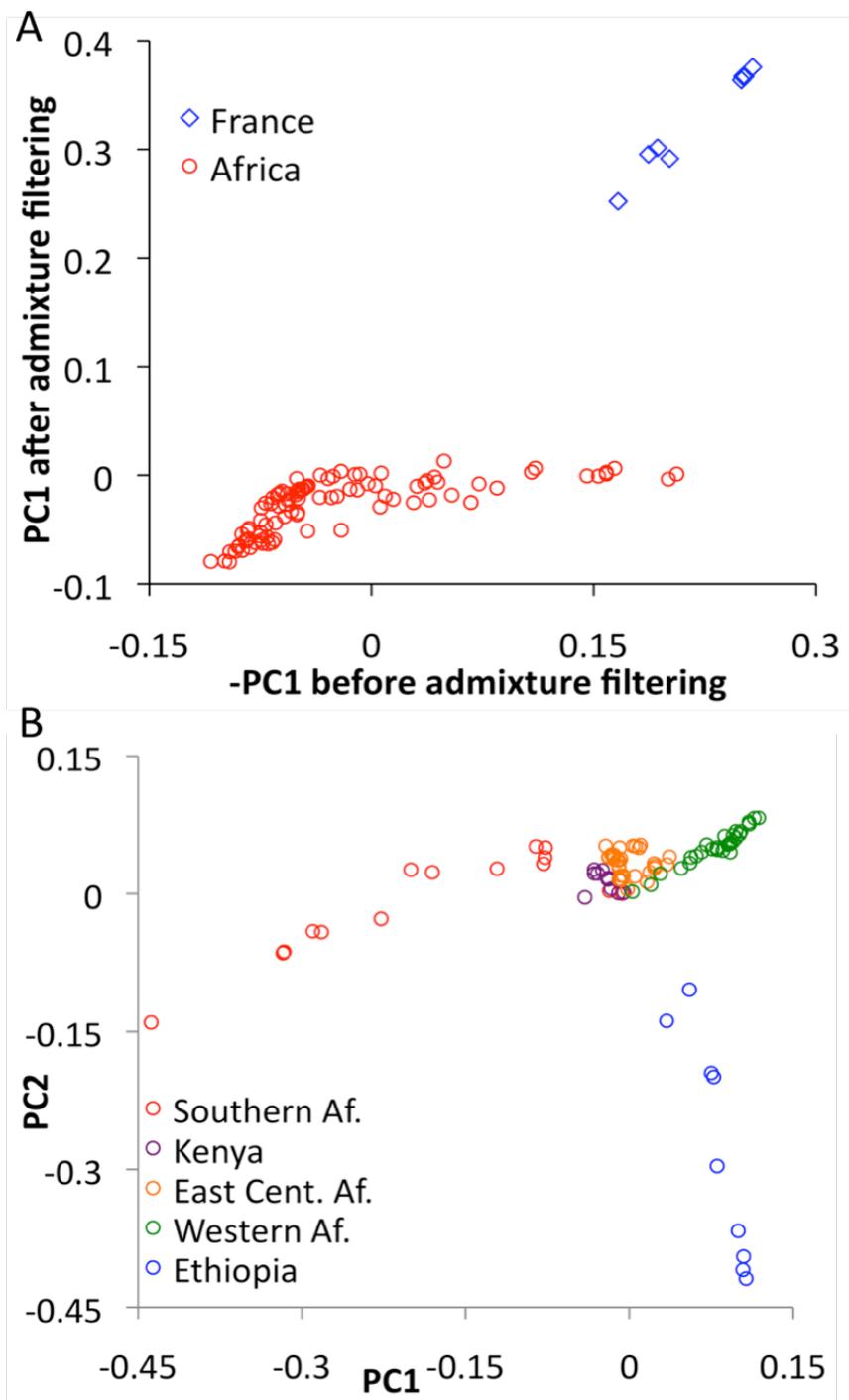

**Figure 5.** Principal Components Analysis (PCA). (A) PCA was done for the full primary core data set before and after masking putative cosmopolitan admixture from sub-Saharan genomes. Reductions in the magnitude of PC1 after filtering are consistent with the admixture identification method being largely successful. (B) PCA was applied to the sub-Saharan genomes only, after admixture filtering. Genomes were found to cluster by geographical region.

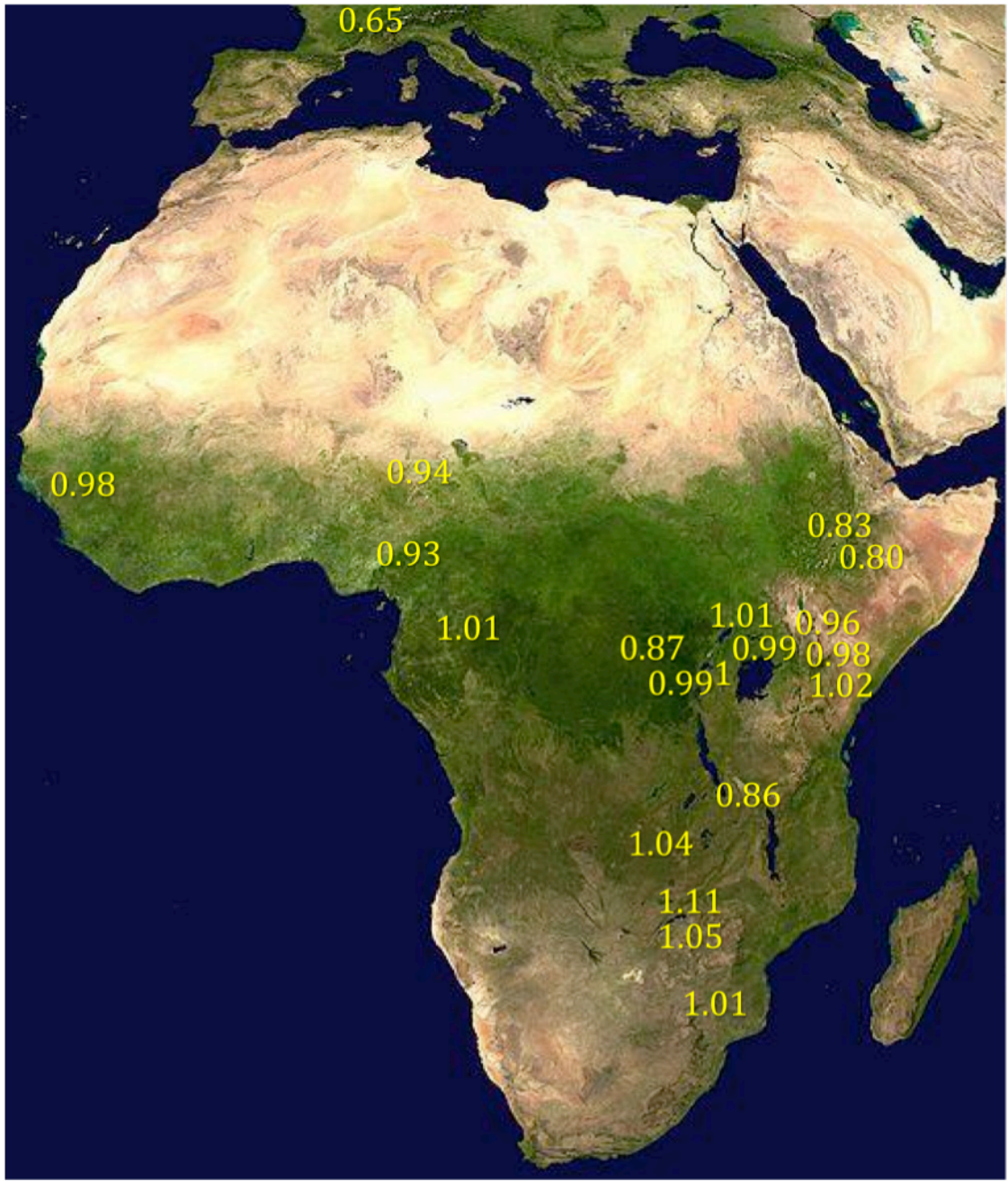

**Figure 6.** Relative nucleotide diversity, scaled by $\pi_{RG}$, was calculated for each population sample. This method allowed the comparison of diversity between populations with missing data in different genomic regions, and allowed the inclusion of secondary core genomes. Values were corrected for the modest predicted effects of sequencing depth (see Materials and Methods), and were based on non-centromeric, non-telomeric chromosomal regions, and equal weighting of chromosome arms.

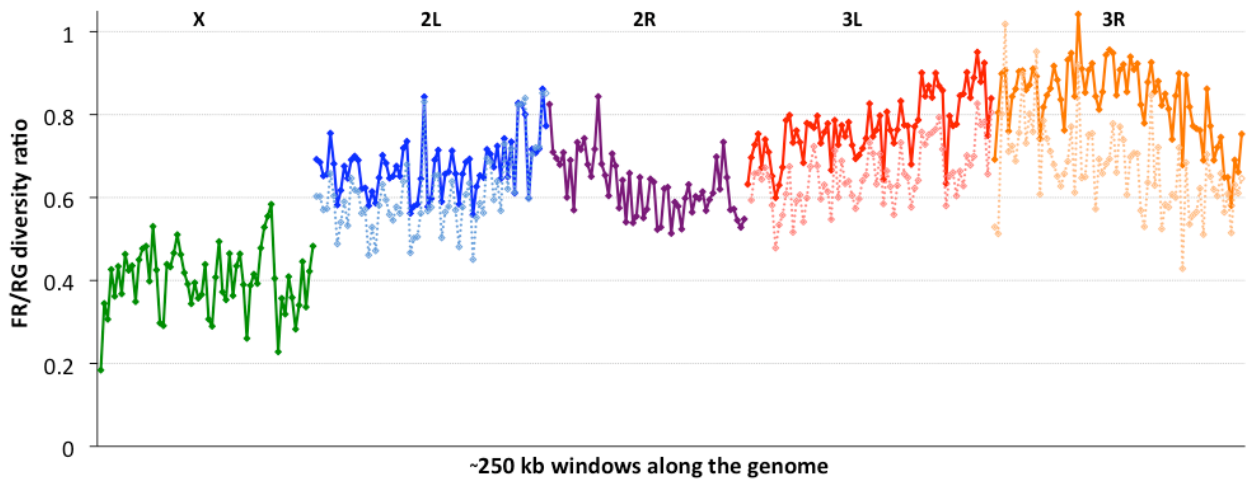

**Figure 7**. The ratio of nucleotide diversity between non-African (France, FR) and African (Rwanda, RG) genomes. Each window contains 5000 RG non-singleton SNPs. Chromosome arms are labeled and indicated by color. Dashed series for the three arms with segregating inversions in the FR sample reflect diversity ratios for standard chromosomes only, indicating that inversions add significant diversity at the scale of whole chromosome arms.

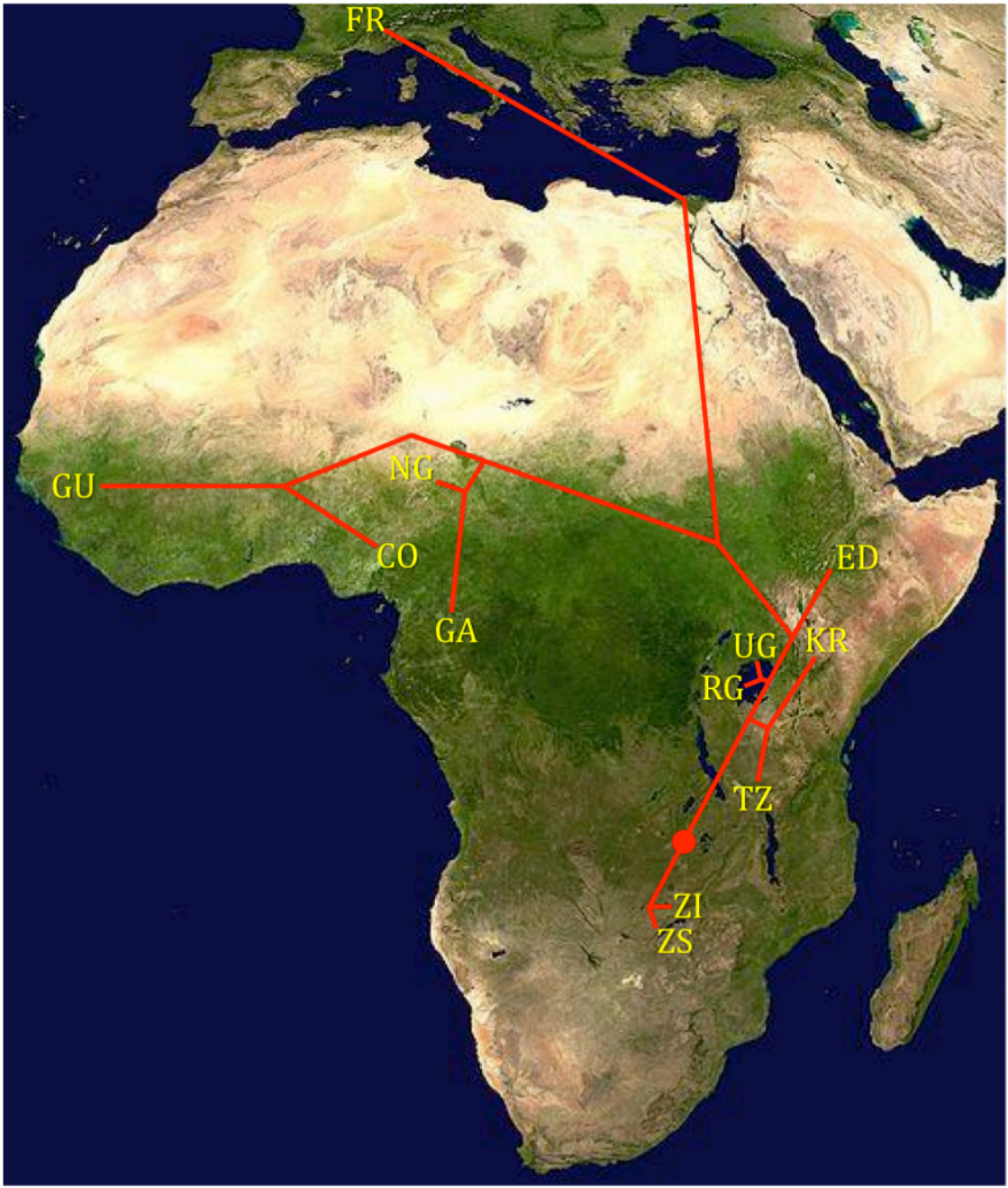

**Figure 8.** Topology of a neighbor-joining population distance tree based on the matrix of $D_{xy}$ values (Table 2). Red dot indicates root based on midpoint rooting. Branch lengths are not to scale.

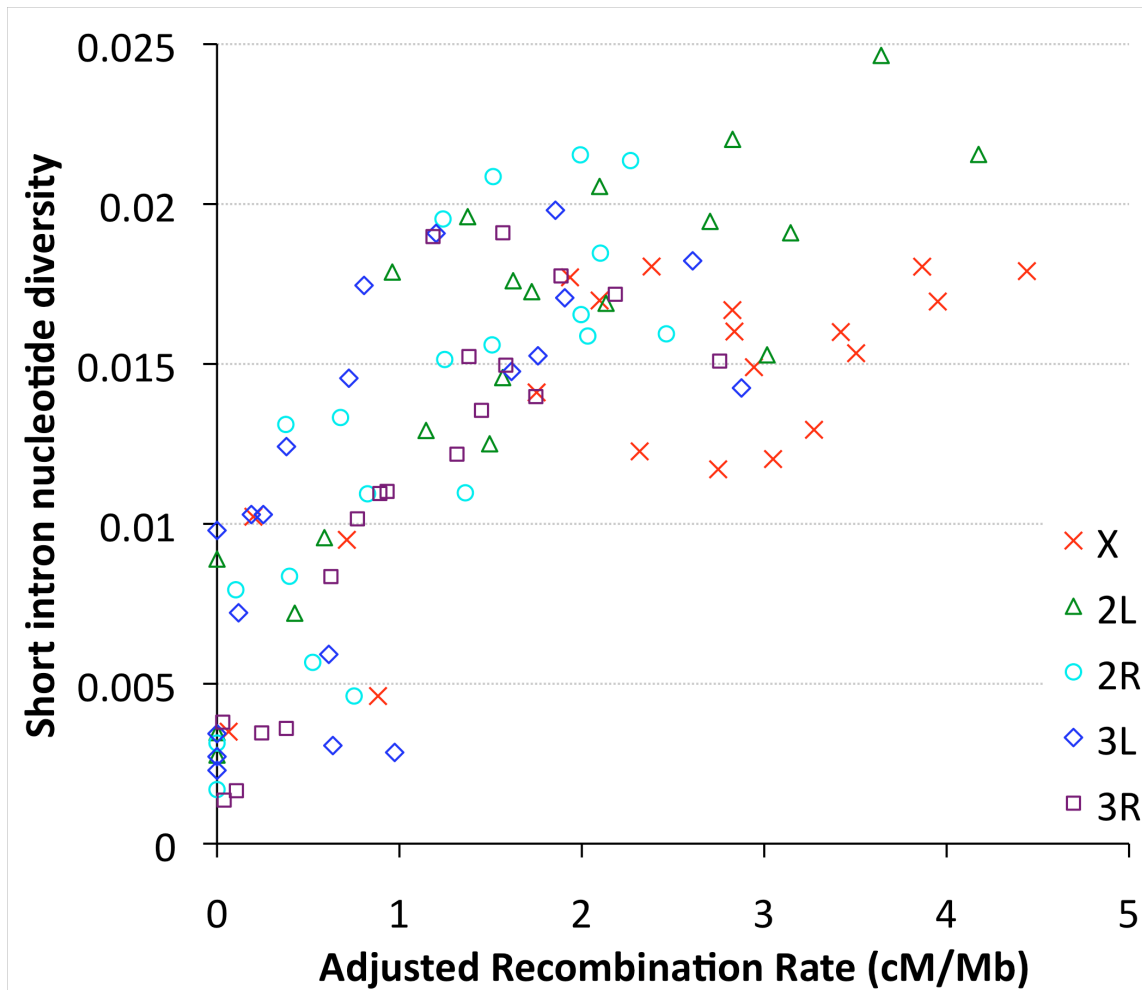

**Figure 9.** Nucleotide diversity vs. recombination rate for short intron sites (bp 8-30 in <65bp introns) is plotted by cytological band. Recombination rate estimates are from Langley et al. (2011), multiplied by one half for autosomes and two thirds for the X chromosome, and weighted by cytological sub-band recombination rate estimates and site counts.

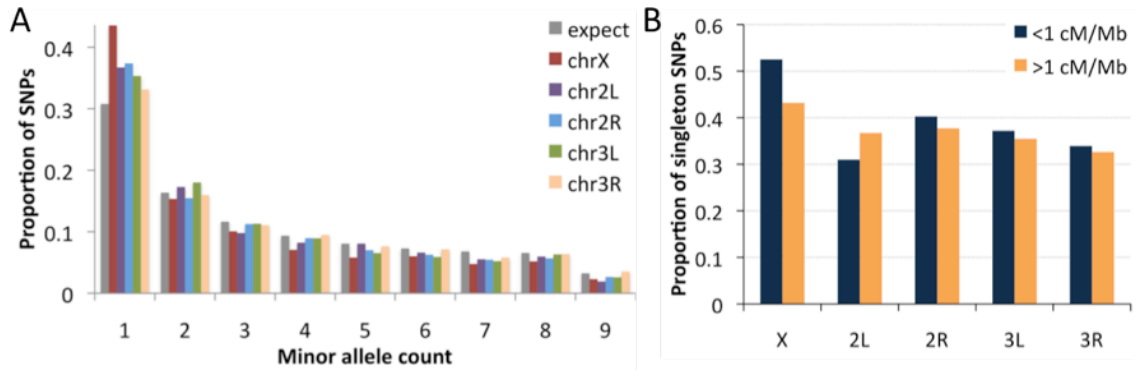

**Figure 10.** Allele frequencies for the RG sample (using a sample size of 18) at short intron sites. (A) The folded frequency spectrum for each chromosome arm. (B) Comparison of the proportion of SNPs with a minor allele count of 1 in regions of lower versus higher recombination.

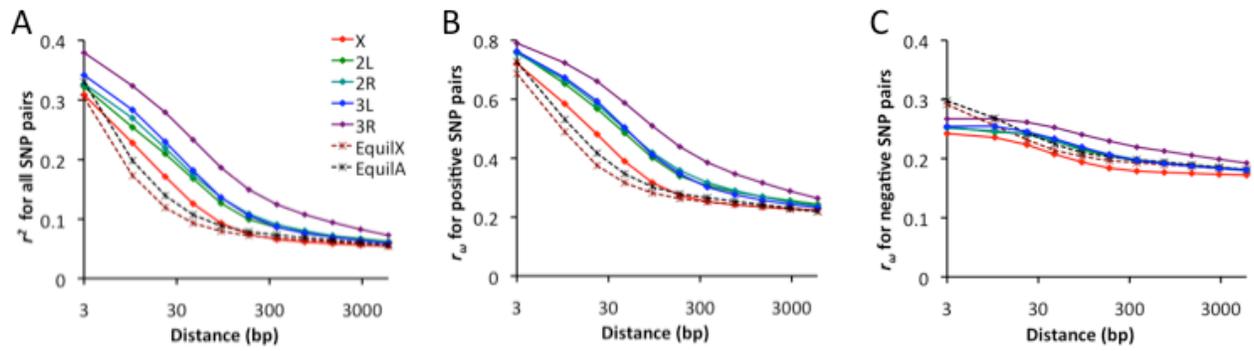

**Figure 11.** Linkage disequilibrium (LD), excluding singleton polymorphisms. Series refer to the observed LD for each major chromosome arm, and the expected LD from neutral equilibrium simulations for X-linked and autosomal loci, as given in panel A. (A) Average $r^2$ for a series of SNP pair distance bins. (B) Average $r_\omega$ for SNP pairs with positive LD. (C) Average $r_\omega$ for SNP pairs with negative LD.